\documentclass[final,3p,times,twocolumn,amsmath,amssymb]{elsarticle}
\usepackage{graphicx}
\usepackage{dcolumn}
\usepackage{bm}
\usepackage{psfrag}

\begin{document}
\begin{frontmatter}

\title{Stripes in cuprate superconductors: Excitations and dynamic
  dichotomy}

\author{G. Seibold\corref{cor1}}
\ead{goetz@physik.tu-cottbus.de}
\cortext[cor1]{corresponding author}
\address{Institut f\"ur Physik, BTU Cottbus, PBox 101344, 03013 Cottbus,
Germany}                                                                            
\author{M. Grilli}
\author{J. Lorenzana}
\address{SMC-INFM-CNR and Dipartimento di Fisica,
Universit\`a di Roma ``La Sapienza", P.le Aldo Moro 5, I-00185 Roma, Italy}

\begin{abstract}
We present a short account of the present  experimental situation of stripes in
cuprates followed by a review of our present understanding of their
ground state and excited state properties. Collective modes, the
dynamical structure factor, and the optical conductivity of stripes are computed
using the time-dependent Gutzwiller approximation applied to realistic
one band and three band Hubbard models, and are found to be in excellent
agreement with experiment. On the other hand, experiments like angle-resolved
photoemission and scanning tunneling microscopy show the coexistence
of stripes at high energies with Fermi liquid quasiparticles at low
energies. We show that a phenomenological model going beyond
mean-field can reconcile this dynamic dichotomy.  
\end{abstract}

\end{frontmatter}

\section{Introduction}
Since the discovery of high-temperature superconductors (HTSC) by Bednorz and
M\"uller \cite{bemu} numerous experiments have evidenced the existence  
of electronic inhomogeneities in these compounds (cf. e.g. Ref. 
\cite{ps2}).
While early on these inhomogeneities where believed to be predominantly    
chemically driven, e.g. due to material imperfections like disorder 
induced by the dopant ions,
it was subsequently realized that the strongly correlated character        
of the cuprate superconductors and thus the electronic subsystem           
itself can favor the formation of inhomogeneities on the nano-scale.
According to the analyis by the Rome group \cite{cast95} the 
inherent reduction of the quasiparticle kinetic energy together
with a short range attractive (e.g. electron-lattice) 
interaction can induce an instability towards
phase separation since competing nesting instabilities are suppressed
due to a residual repulsion of the quasiparticles at large momenta.
On the other hand, the long-range
repulsive Coulomb interaction will spoil the associated zero-momentum
instability in the charge sector and instead shift the wave-vector
of the ordering transition to finite values. This so-called frustrated
phase separation mechanism \cite{FPS} thus results in an 
incommensurate charge ordering (CO). 

An alternative approach is based on Hartree-Fock (HF) investigations 
of Hubbard (and tJ)-type hamiltonians \cite{zaa89,mac89,hsch90,pol89}
which have revealed solutions with a combined charge- and 
spin-density wave. In contrast to the frustrated phase separation
mechanism the charge order in this case is driven by a spin-density
wave instability which via the coupling between longitudinal spin
and charge degrees of freedom results in a concomitant charge
density wave. These so-called stripe solutions have 
been confirmed later on by more sophisticated numerical methods. 
Within a density matrix renormalization group (DMRG) approach White and
Scalapino \cite{WS,WS2,WS3} have found stable domain wall solutions 
in the physically relevant doping regime of the tJ-model.
The stripe stability in the tJ-model has also been investigated using
exact diagonalization \cite{HELLBERG,TOHYAMA} as well as quantum and
variational Monte Carlo techniques \cite{BECCA,HIMEDA}. 
Besides the HF approach a variety of methods has also been applied
to Hubbard-type models in order to investigate the possibility of 
charge and spin order. Fleck and collaborators \cite{fle01b} have 
found stable stripes using dynamical mean-field theory (DMFT) and
a cluster perturbation approach has been applied in Ref. \cite{ZACHER}. 
From an ab-initio perspective stripe order in lanthanum cuprates has also been 
shown to be stable in LDA+U calculations \cite{anisimov04}.

Here we present a short update of the current experimental situation (Sec.~\ref{exp}) followed by an overview over
static and dynamical properties of stripes\cite{sei981,sei982,lor02,lor03,sei04,sei05,sei06,sei07,sei09,sei11}
obtained within the time-dependent Gutzwiller approximation (TDGA)\cite{sei01,goe03,goe042,goe08} 
and phenomenological models.  
In Sec. \ref{sec1} we first derive   
the parameter set for the extended Hubbard model which in the following
sections allows us to make quantitative comparison with 
experimental data. We then show in Sect. \ref{sec2} that the Gutzwiller
approximation (GA) of this  model in fact 
leads to stable stripe ground states which allows for the
doping dependent evaluation of the corresponding incommensurate spin response.
In Sec. \ref{sec3} we discuss the electronic structure and various
transport properties of stripes. Sec. \ref{sec4} focusses on the
momentum and frequency dependent spin excitations where we also address
recent resonant inelastic x-ray scattering experiments. 
Sec. \ref{sec5} presents results for the charge excitations from
stripe ground states, focussing on the optical conductivity. Finally, we
show in Sec. \ref{sec6} how our Gutzwiller variational theory can be 
extended towards a dynamical description of stripes before we 
summarize our discussion in Sec. \ref{sec7}.

In the following section we start with an update of the experimental
situation. 
\section{Brief Experimental Review}
\label{exp}
First evidence for stripe order in HTSC's
came from elastic neutron scattering experiments by Tranquada
and collaborators \cite{tra95,tra96,tra97}.
They observed a splitting of both spin and charge order 
peaks in La$_{1.48}$Nd$_{0.4}$Sr$_{0.12}$CuO$_4$ (LNSCO)
which resembled similar data in the nickelates
where both incommensurate antiferromagnetic (AF) order
\cite{nio1,nio3} and the ordering of charges \cite{nio2,nio3} has been
detected by neutron scattering and electron diffraction, respectively.
In Ref. \cite{nio3} it was shown that the magnetic ordering 
displays itself as an occurence
of first and third harmonic Bragg peaks whereas the charge ordering
is associated with second harmonic peaks. 
Meanwhile analogous static stripe order
has also been detected in Eu -codoped lanthanum cuprates (LECO) \cite{klauss00} 
and La$_{2-x}$Ba$_{x}$CuO$_4$ (LBCO) \cite{fujita021,fujita04} which all display
a LTT lattice distortion. That the latter is not necessarily a condition
precedent to the formation of stripes has recently been demonstrated 
by applying pressure to a La$_{1.875}$Ba$_{0.125}$CuO$_4$
sample. This restores the structure to fourfold lattice symmetry
keeping symmetry broken electronic stripe states\cite{huecker10}.
Moreover, charge and spin stripe order can also be induced in the LTO
phase of impurity (Cu, Zn, Fe, Ga) substituted lanthanum cuprates 
\cite{fujita09} without the need of applying pressure.
Whereas charge order in the neutron and hard x-ray scattering 
\cite{zimmermann} experiments mentioned above has only been
detected indirectly via the associated lattice modulation, more recent
soft resonant x-ray scattering studies on LBCO \cite{abb05}
and LECO \cite{fink09} have directly revealed the spatial 
modulation of charge in these systems.

The incommensurability $\delta$ 
(defined as the shift of the magnetic Bragg peaks 
from the antiferromagnetic (AF) wave-vector $Q_{AF}$ 
in reciprocal lattice units) in  samples with long range stripe order
follows the so-called 'Yamada-Plot' \cite{yam98} 
and depends linearly on doping $\delta=n_h$ up to $n_h \approx
1/8$. Here $n_h$ is the number of
added holes per planar Cu with respect to the parent insulating compound.
This behavior is compatible with stripe like modulations of charge and  spin
having a linear concentration of added holes $\nu=1/2$ (so-called
half-filled stripes) as we will discuss in more detail in Sec.~\ref{sec2}.

Evidence for some kind of stripe order in other HTSC materials is 
based on the doping dependence of the low energy spin response.
In non-codoped lanthanum cuprates (LCO) the incommensurability $\delta$
(defined in terms of the shift of the low energy magnetic scattering 
from $Q_{AF}$) also follows the 'Yamada-Plot' of the samples with
long-range order. This suggests that the static stripe order  
is replaced by some kind of 'fluctuating order' in
the Nd- (or Ba-, Eu-) free samples.
Above $n_h\approx 1/8$, the incommensurability
stays essentially constant but the intensity
of the low energy spin fluctuations decreases and vanishes at the same       
concentration where superconductivity disappears in the overdoped
regime \cite{waki04}. 
This behavior is mirrored at low doping in that superconductivity, in
LSCO materials, disappears upon decreasing doping exactly at the point in which incommensurate
scattering parallel to the CuO bond disappears.\cite{fujita02}
This strong correlation between superconductivity and low energy 
incommensurate scattering parallel to the CuO bond      
suggesting an intimate relation between both phenomena and provides a
strong motivation to understand stripe physics in cuprates.

Once superconductivity disappears by decreasing doping 
 below $n_h\approx 0.055$ in lanthanum cuprates 
incommensurate scattering does not disappear entirely 
but rotates by $45^0$ \cite{waki99,waki00,mats00,fujita02}
to the diagonal direction.  
Additionally the orthorhombic lattice distortion
allows one to conclude that
the elastic diagonal magnetic scattering is one-dimensional with the associated
modulation along the orthorhombic $b^*$-axis, supporting again the picture
of stripe formation.
When $\delta$ is measured in units of  reciprocal tetragonal
lattice units in both the vertical and diagonal phase it turns out that
the magnitude of the incommensurability numerically coincides across
the rotation leading to a linear relation  $\delta=n_h$.
Upon approaching the border of the AF phase at $n_h=0.02$
the incommensurability
approaches $\epsilon=n_h$ \cite{matsuda02} where $\epsilon$ is
measured in units of
reciprocal orthorhombic lattice units, thus $\epsilon=\sqrt{2}\delta$.
The crossover from diagonal to vertical
incommensurate spin response is so far only observed in lanthanum cuprates
whereas in all other HTSC compounds the scattering is always along the
Cu-O bond direction. Despite this peculiarity of LCO other experiments point to universal behavior. 

A linear relationship between doping and incommensurability is 
also observed in YBCO$_{6+x}$ although the curve falls below the one 
of lanthanum cuprates \cite{arai99,dai01,haug10}. Apart from details,
this suggests that 
stripe physics is a universal phenomenon in cuprates. 

Upon increasing frequency of spin excitations the dispersion develops
a spectrum ('hour glass')  which is similar in many high-T$_c$
cuprates again suggesting universality.  
At a certain
frequency $E_s$ the incommensurate branches merge \cite{note1} 
at the AF wave-vector
and for even higher energies the excitations become incommensurate 
again. The 'hour glass' spectrum has been detected in the vertical
\cite{tra04,chr04,vignolle,kofu07} and diagonal phase 
\cite{matsuda08,matsuda11} of lanthanum cuprates, the YBCO compounds 
\cite{hayden04,reznik04,stock05,hinkov07,hinkov08} and 
Bi$_2$Sr$_2$CaCu$_2$O$_{8+y}$ \cite{fauque07,xu09}.

Magnetic fluctuations which dispersion follows the 'hour-glass' 
shape naturally arise from stripe correlations in the 
ground state 
\cite{sei05,sei06,voj04,uhr04,uhr05,moe04,vojta06,and07,and10,kru03,car04,car06,sei09} (cf. Sec. \ref{sec4}).
Basically a Goldstone mode arises from each of the incommensurate
wave-vectors and disperses in a cone-shaped structure to higher energies.
However, due to disorder and peculiar magnetic couplings across the stripes 
(as supported
by recent inelastic neutron scattering (INS) experiments on La$_{2-x}$Sr$_x$CoO$_4$ \cite{boothroyd11}) 
the weight of the outwards dispersing contribution is strongly 
suppressed so that
the excitations which dominate the spectral weight disperse towards
the AF wave-vector $Q_{AF}$. The corresponding excitation energy $E_s$ 
at $Q_{AF}$ is  
essentially determined by the spin coupling across the antiphase domains.
Since at high energies the excitations should again resemble those of
the (undoped) AF one finds an outwards dispersing intensity also above
$E_s$. 

While apparently the spectrum of spin excitations 
is compatible with stripe correlations the problem is the 
detection of an associated charge order in the non-codoped
lanthanum cuprates, YBCO and bimuthates. Experimental
data of local probes like NQR \cite{kraem99,tei00,sing02},
NMR \cite{has02}, EXAFS \cite{BIANCONI}, and 
X-ray microdiffraction \cite{FRATINI}
 are suggestive of the formation of electronic
inhomogeneities, possibly induced by magnetic field \cite{julien11}, 
but cannot provide evidence for long-range order.
In this context it is remarkable that 
due to refinements in the experimental technique Haase et al. \cite{has02} 
where able to demonstrate a correlation of charge and
density variations on short length scales.

One way to reconcile the 'stripe-like' magnetic excitation spectrum
with the absence of long-range charge order is the notion of a charge
nematic \cite{vojta} which (in the charge channel) breaks rotational
but not translational invariance. This concept is extremely useful
in detwinned YBCO where both neutron scattering \cite{hinkov07,hinkov08} and
thermoelectric transport \cite{daou10} data show pronounced in-plane 
anisotropies
but may also apply to the spin-glass phase of lanthanum cuprates where,
as mentioned above, the incommensurate spin scattering is one-dimensional.

Due to the absence of clear signatures of charge ordering in other
than co-doped lanthanum cuprates alternative theories have
been proposed in order to account for the incommensurate
spin scattering and the associated 'hour glass' magnetic
spectrum \cite{brinckmann99,norman00,chubu01,schnyder04,berciu04,sherman04,eremin05,sherman11,yamase06}. Nonetheless there are a number of 
fingerprints in spectroscopic probes which support the presence of 
incommensurate charge scattering.
For example, the anomalous lineshape and temperature dependence of the
bond-stretching phonons in various cuprate materials 
(cf. Ref. \cite{reznik10} and references therein) 
can be attributed to the influence of stripes.
For example, it has been proposed that the softening of these
phonons at certain momenta is due to the influence of charge stripe 
fluctuations on the phonon self-energy \cite{kane02}  In addition
one may also have a Kohn-type anomaly due to the nesting
along the half-filled stripe \cite{mukhin07}.
Furtheron, the scattering of charge carriers by incommensurate charge 
fluctuations should induce a reconstruction of the Fermi surface 
\cite{ZACHER,salkola,sei00}.
Angle-resolved photoemission experiments (ARPES) on LSCO \cite{zho99,zho01} 
have indeed shown
the appearance of straight Fermi surface (FS) segments around the M-points
of the Brillouin zone as expected for a striped ground state. However, 
one peculiar feature of these experiments concerns the fact that the
stripe-like FS was obtained from the momentum distribution
$n_{\bf k}$, by integrating
the spectral function over a broad energy window ($\sim 300$
meV). On the other hand, upon following the momentum dependence 
of the low-energy part of the energy distribution curves 
a large FS was found corresponding to the LDA band-structure 
and fulfilling Luttinger's theorem.
We will address this dichotomy of stripe-like spectral
features at  large energies and  a 'protected' FS at
low energies in Sec. \ref{sec6}.

In LECO, where static charge order has been detected with RIXS \cite{fink09},
the associated reconstruction of the FS has indeed been resolved by ARPES.
Moreover, a recent series of magnetotransport experiments 
\cite{doiron1,yelland,bangura,leboeuf} have revealed quantum oscillations
suggestive of the formation of electron pocktes due to a FS reconstruction.
It has been shown by Millis and Norman \cite{millis07} 
that a charge density wave ground state
with  intermediate values of the stripe order parameter is in qualitative
agreement with the quantum oscillation data.  A recent comparative study
of thermoelectric properties in YBCO and LECO \cite{laliberte11} also 
 provides strong evidence that the Fermi surface reconstruction in YBCO
is associated with stripe correlations.
In the context of transport experiments one should also mention the Nernst
effect which in hole doped HTSC's above T$_c$ 
usually yields a large positive signal \cite{wang06}
and which has been considered as indicative of fluctuating superconductivity.
However, in addition also a FS reconstruction contributes to the
Nernst signal and both contributions have been resolved
in LNCO and LECO \cite{choin09}. These authors have attributed the
high temperature signal to originate from changes in the FS, an interpretation
which has recently been challenged in Ref. \cite{hess10}.
A theoretical analysis of the Nernst effect in terms of stripe order
can be found in Refs. \cite{martin10, hackl10}.

Scanning tunneling microscopy (STM) provides additional evidence
(limited obviously to the surface) for charge order  in cuprate
superconductors. Corresponding measurements performed on bismuthate and       
oxychloride superconductors have revealed a complex modulation              
of the local density of states (LDOS) both in the superconducting (SC)      
state \cite{hoff02,how03,elroy05,hashi06,hana07,wise08} and above $T_c$     
\cite{wise08,versh04,hana04}. In both cases one observes peaks in           
the Fourier transform of the                                                
real space LDOS at wave-vectors $Q=2\pi/(4a_0)...2\pi/(5a_0)$ suggestive of 
checkerboard or stripe charge order. However, it is important to
distinguish between the case where these peaks are non-dispersive in 
energy (and thus signature 
of 'real' charge order) or the situation where they follow a bias-dependent 
dispersion due to quasiparticle (QP) interference. 
In the latter case the spatial LDOS variations can be understood                
from the so-called octet model \cite{hoff02,wang03}                             
which attributes the modulations to the                                         
elastic scattering between the high density regions of the                      
Bogoljubov 'bananas' in the superconducting state.                              
Recent STM investigations \cite{kohsaka08,alldredge08} may resolve this         
apparent conflict since they suggest that both, dispersive and non-dispersive   
scattering originates from different regions in momentum and energy space.      
The states in the nodal region                                                  
which are well defined in k-space and undergo a transition                      
to a $d$-wave SC state below $T_c$ are then responsible for the low energy      
QP interference structure of the LDOS, whereas the ill-defined 
k-space 'quasiparticle' states in the antinodal regions 
are responsible for the non-dispersive CO above some energy scale $\omega_0$.
As mentioned before we will address the issue of high energy electronic
order in Sec. \ref{sec6}.

\section{Model and parameters}\label{sec1}

Moment formation in a spin-density wave solution involves the 
large Hubbard $U$ scale and modifies 
the electronic structure of the system over a large energy range.
Therefore, it is not {\it a priori} obvious that the reduction
of a multiband model to a more simplified model hamiltonian
will retain the relevant features in the spectra, 
especially when one is concerned with
high frequency excitations. A suitable method in order to
deal with this problem has to fulfill essentially the following
two conflicting criteria: {\it (a)} it should be sufficiently accurate 
in order to account for the Hubbard-type correlations of the model and
{\it (b)} it should allow for the investigations of large systems
in order to provide a sufficient resolution for the correlation functions
in momentum space. 

Subsequent to the first stripe calculations 
\cite{zaa89,mac89,hsch90,pol89}, investigations
of collective excitations have
been performed within the Hartree-Fock (HF) approximation of the one-band
Hubbard model supplemented with RPA fluctuations 
\cite{ichi01,kan01,kan02,kan022,varl02}.
However, in case of the HF approximation\cite{inu91} 
it turns out that the stripe solutions relevant for cuprates
are only favored for unrealistic small
values of $U/t\approx 3 ... 5$ whereas a ratio of $U/t\approx 8$
is required to reproduce the magnon spectrum of LCO (cf. below).

In order to enable a quantitative comparison with experiment,
the results we will discuss in the following sections 
will be mostly based on
the Gutzwiller approximation (GA) \cite{gut65} supplemented with Gaussian
fluctuations \cite{sei01,goe03,goe042,goe08} (so-called time-dependent
GA [TDGA]).

\begin{figure}[htb]
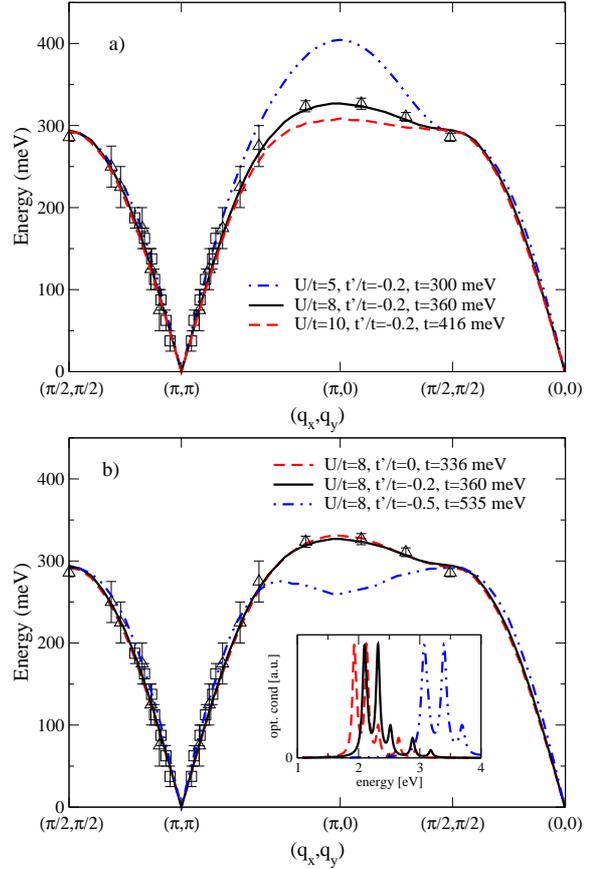

\includegraphics[width=7.5cm,clip=true]{fig1a.eps}
\includegraphics[width=7.5cm,clip=true]{fig1b.eps}
\caption{Energy and wave vector dependence of magnetic
excitations in the half-filled system as obtained from
the TDGA for (a) $t'/t=-0.2$ and varying $U/t$;
(b) $U/t=8$ and varying $t'/t$. Squares
and triangles correspond to data points from INS experiments on LaCuO$_4$
by Coldea et al. \protect\cite{col01}. The inset to (b) shows the optical
conductivity for the same parameters as used in the main panel (b).}
\label{fig:edq}
\end{figure}

We consider the one-band Hubbard model
\begin{equation}\label{HM}
H=\sum_{i,j,\sigma} t_{ij} c_{i,\sigma}^{\dagger}c_{j,\sigma} 
+ U\sum_{i}n_{i,\uparrow}n_{i,\downarrow},                                           \end{equation}
where $c_{i,\sigma}$ ($c^\dagger_{i,\sigma}$) destroys (creates) an electron 
with spin $\sigma$ at site $i$, and $n_{i,\sigma}=c_{i,\sigma}^{\dagger}c_{i,\sigma}$.
$U$ is the on-site Hubbard repulsion and $t_{ij}$ denotes the hopping 
parameter between sites $i$ and $j$. We restrict to hopping 
between nearest ($\sim t$) and next-nearest ($\sim t'$)neighbors.  
For high-T$_c$ cuprates the value of $U$ can be estimated from the
magnon dispersion of the undoped system \cite{col01,headings10}.
In fact, within spin-wave theory (SWT) applied to the Heisenberg model 
(corresponding to $U/t \rightarrow \infty$) the magnon excitations are
given by
\begin{equation}
\hbar\omega_{\bm q }\propto \sqrt{1-[\cos q_x +\cos q_y]^2/4}
\end{equation}
and thus the dispersion vanishes along the magnetic Brillouin zone.
Including quantum fluctuations beyond SWT \cite{singh95} leads to excitations
with lower energy at $(\pi,0)$ than at $(\pi/2,\pi/2)$, contrary
to what is observed in undoped cuprate superconductors \cite{col01,headings10}.
It has been argued that in
cuprates corrections to the Heisenberg model arising as higher orders
in a $t/U$ expansion are relevant \cite{rog89,sch90,lem97,lor99,col01}.
The most important of such corrections is a term which cyclically
exchanges four spins on a plaquette. A sizable value for this term
has been revealed by analyzing phonon-assisted multimagnon infrared
absorption\cite{lor99} and the dispersion
relation measured with INS\cite{col01,headings10} shown in
Fig.~\ref{fig:edq}.  In particular,
the dispersion in the $(\pi,0)$ and $(\pi/2,\pi/2)$ is mainly due
to this term. 
In case of the Hubbard model the magnetic zone edge dispersion is 
due to the finiteness
of $U/t$. For moderate values of $U/t$ this leads to dispersion
which decreases from $(\pi,0)$ to $(\pi/2,\pi/2)$  along the magnetic
zone boundary whereas for large $U/t$  the behavior of the spin-1/2 
Heisenberg model is recovered\cite{zheng05}.

Fig. \ref{fig:edq}a shows the $U/t$ dependence of the magnetic
excitations for the half-filled AF as compared to INS data from 
Ref. \cite{col01}. The dispersion along the magnetic zone boundary
can be accurately fitted by a value of $U/t=8$ where the
absolute energy scale is fixed by the nearest neighbor hopping
$t=360 meV$. Panel b of Fig. \ref{fig:edq} displays the dependence
of the spin excitations on the next-nearest neighbor hopping $t'/t$.
For small to moderate values of $t'/t$ the zone boundary dispersion
is only slightly dependent on this parameter, however, significant softening
occurs for $t'/t=-0.5$. The increasing frustration between nearest and
next-nearest neighbor magnetic interactions drives the system towards
an 1D-AF instability the precursor of which is seen as 
a magnetic mode softening at ${\bf q}=(\pi,0)$. 

Further information
on the parameter $t'/t$ can be obtained from the optical conductivity
which in the inset to Fig. \ref{fig:edq}b is shown for the same
parameters as used in the main panel.
The lowest energy excitation $\Omega_{min}$ for the half-filled system 
is shifted up with increasing $|t'/t|$ and the experimental value of 
$\Omega_{min} \approx 2.1 eV$ for LCO \cite{uch91,oka11} is accurately
reproduced for the parameter set $U/t=8$, $t'/t=-0.2$, $t=360 meV$.
In the following section we
will show that a value of $t'/t=-0.2$ is also appropriate to account for
the doping
dependent incommensurability of lanthanum cuprates \cite{yam98} 
based on a striped ground state. 

Note that some few results discussed in the present review 
(Secs. \ref{sec2a}, \ref{sec5}) are also
obtained within the three-band model
which for copper $d_{x^2-y^2}$ and 
oxygen $p_{x,y}$ orbitals includes the associated hopping processes, 
orbital energies, 
and intra- and interorbital Coulomb repulsions. 
We have taken these parameters either from              
first principle computations \cite{mcm90,hybert} or from more
empirical considerations \cite{esk90}.

\section{Stripe stability and static properties}\label{sec2}
\subsection{Charge and spin structure}\label{sec2a}
\begin{figure}[htp]
\includegraphics[width=7.5cm,clip=true]{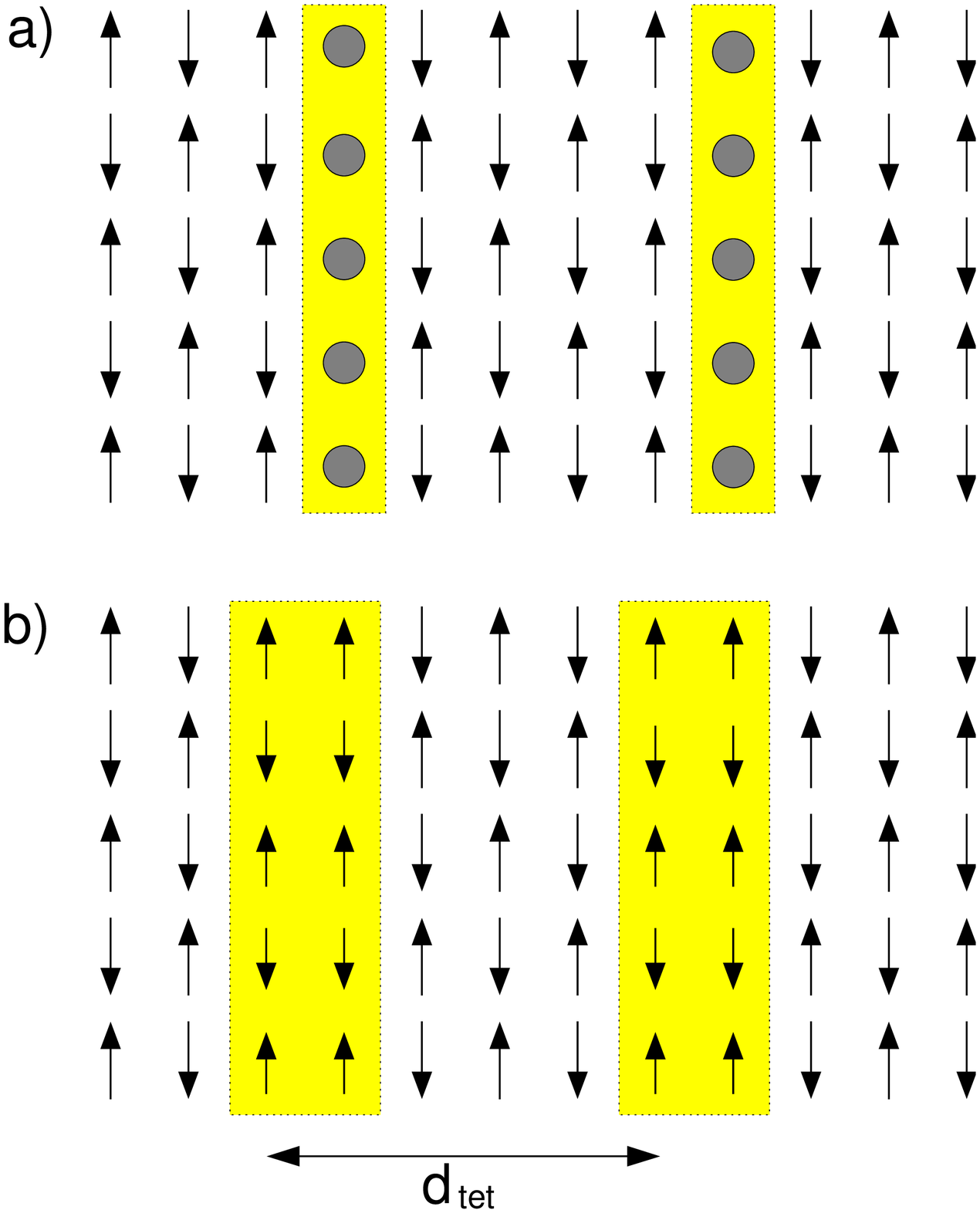}

\vspace*{0.3cm}

\includegraphics[width=8cm,clip=true]{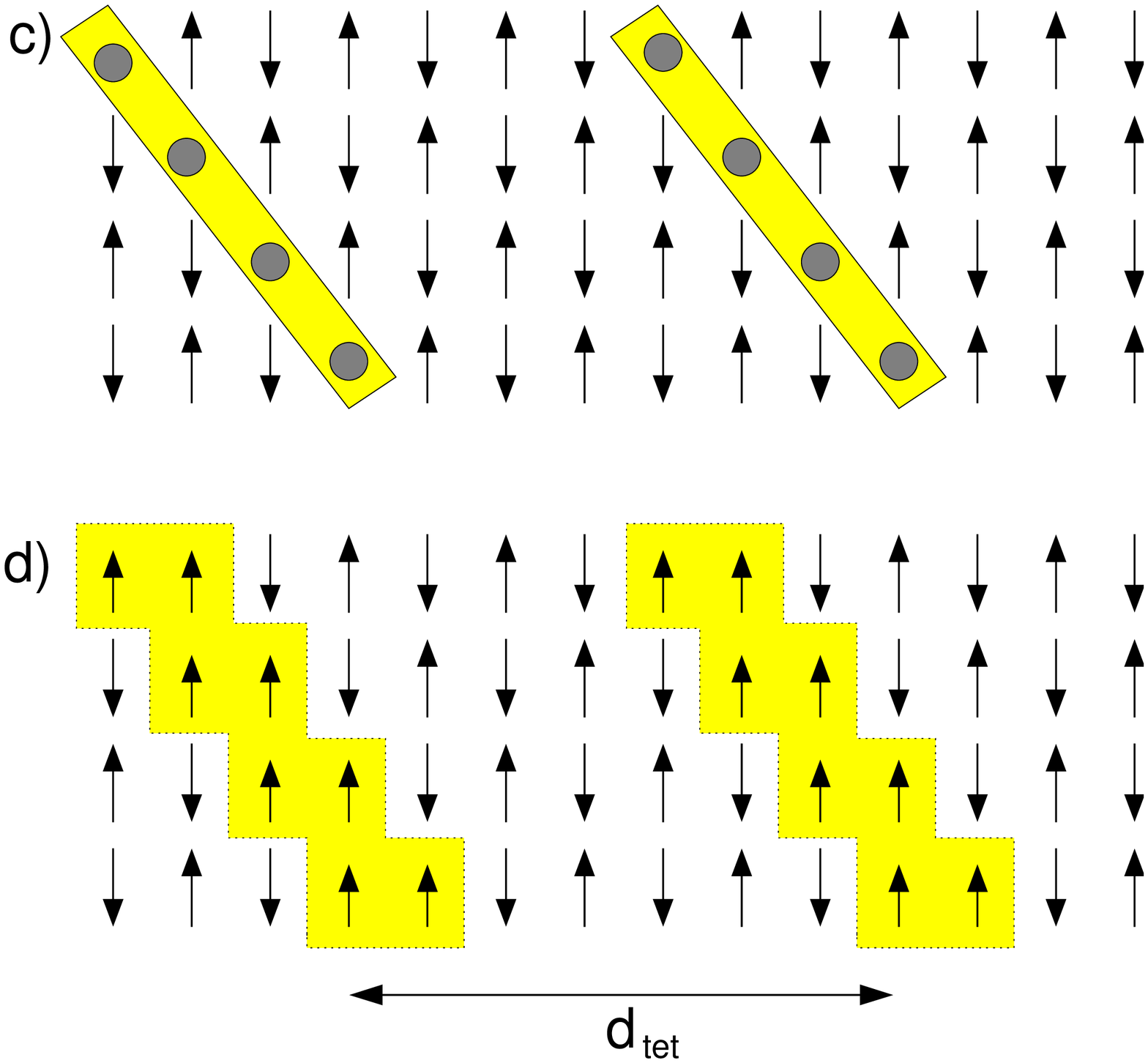}
\caption{Spin structure of vertical (a,b) and diagonal stripe textures.
The doped holes mainly reside on the (shaded) domain walls.
In both cases the prototypical structures comprise site-centered (a,c)
and bond-centered structures (b,d). The distance between charge stripes,
measured along the copper-oxygen bond direction is denoted as $d$
in units of the tetragonal lattice constant $a_{tet}$.  }
\label{ststruc}
\end{figure}

Stripes are characterized by one-dimensional 
antiphase boundaries for the AF order which also host
the doped holes. These textures can be understood as the strong coupling remnant of an
instability in the spin channel. It can be shown \cite{marki10}
that for dispersion relations relevant for cuprates
the orientation is either along the copper-oxygen (vertical)
or the diagonal direction. The domain walls can either reside on
lattice sites or bonds and these two prototypical
stripe solutions for both vertical and diagonal orientations 
are sketched in Fig. \ref{ststruc}.
   
By symmetry, for site-centered (SC) stripes the magnetization vanishes
on the domain wall whereas for bond-centered (BC) stripes these
are build up from ferromagnetically aligned nearest-neighbor spins.
For diagonal BC stripes this  leads to the situation that configurations
with $d=even$ acquire a macroscopic magnetization in contrast
to vertical BC stripes where the ferromagnetic bonds alternate along
the domain wall. Diagonal BC stripes can be viewed as a 
realization of the staircase structures proposed by Granath \cite{granath04}
with step length $l=2$.

\begin{figure}[htb]
\begin{center}
\includegraphics[width=8cm,clip=true]{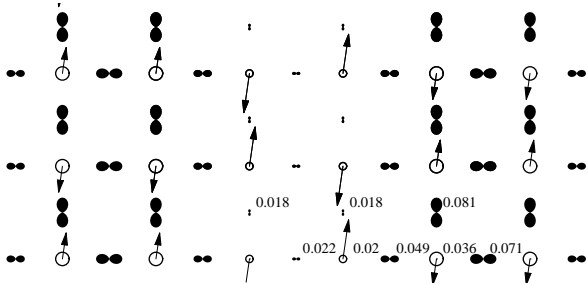}
\caption{Charge and spin density for $d=4$ BC 
vertical stripes. Doping $n_h=1/8$. 
Open circles ($p$ orbitals) represent Cu (O) sites. The
numbers and the size of symbols  represent the excess charge  whereas the
spin density is proportional to the length of arrows.   
}
\label{rhosp}
\end{center}
\end{figure}

Fig.~\ref{rhosp} shows the charge and spin density for $d=4$ BC 
vertical stripes (doping $n_h=1/8$ ) obtained within the 3-band model 
and parameters taken from Ref.~\cite{esk90}. 
A similar charge distribution was predicted in Ref.~\cite{lor02} and
found to be in excellent agreement with a charge sensitive probe by
Abbamonte and collaborators\cite{abb05}. However, these authors 
could not determine if the 
stripes where Cu centered or O centered. More recently Davis
and collaborators\cite{koh07} have imaged glassy stripes which indeed are 
centered on O as  predicted\cite{lor02}.  Taken together these experiments
give us amplitude and phase information of stripes in excellent
agreement with Ref.~\cite{lor02} thus showing that it is possible to obtain 
and even predict realistic information on the intermediate scale 
physics of cuprates. 

\begin{figure}[htb]
\begin{center}
\includegraphics[width=7.5cm,clip=true]{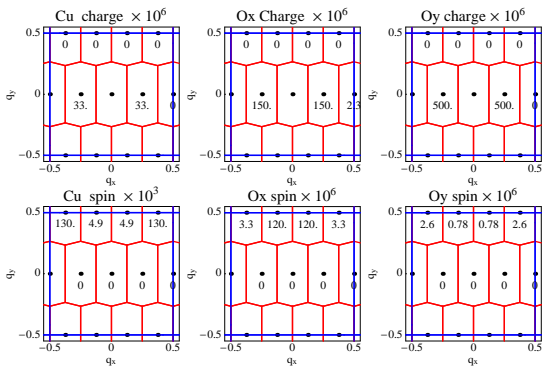}
\caption{Bragg weights $m_{\bm{Q}}^2$, $\rho_{\bm{Q}}^2$ at reciprocal
  lattice vectors (indicated by the dots) for $n_h=1/8$ and $d=4$
  stripes. The polygons around each
  point are the reduced magnetic Brillouin zones. We use reciprocal
  lattice units of the fundamental Brillouin zone. O$_x$ (O$_y$) are the
  oxygens in the bonds perpendicular (parallel) to the stripe. 
}\label{bragg}
\end{center}
\end{figure}

The Fourier transform of the charge ($\propto \rho_{\bm{Q}}$) and 
spin distribution ($\propto m_{\bm{Q}}$) determines
Bragg peak weights ($\propto \rho_{\bm{Q}}^2, m_{\bm{Q}}^2$) 
in scattering experiments
(for the definitions see Ref.~\cite{lor05}) as shown in
Fig.~\ref{bragg}.  Disregarding the 
difference in cross section for different processes (magnetic neutron scattering
vs. X-ray or nuclear neutron scattering) 
we see that Bragg weights for the charge are practically 3 orders of
magnitude smaller than those for the Cu spins. 
This is due to very soft
charge distribution shown in Fig.~\ref{rhosp} with respect to the spin
distribution and explains why it has been so hard to detect  charge
ordering.  

\subsection{Stripe stability and incommensurability}
\begin{figure}[htb]
\includegraphics[width=7.5cm,clip=true]{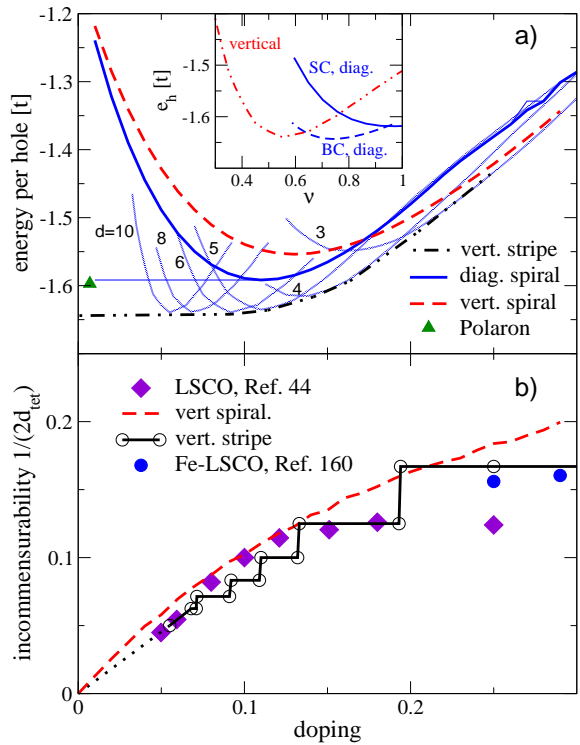}
\caption{(a) Binding energy per hole $e_h$ as a function of doping 
for stripes, uniform spirals and spin polaron.  
The inset shows $e_h$ for $d=10$ vertical (dashed-dotted), 
bond centered diagonal (dashed), and site-centered diagonal (solid) stripes
as a function of the filling factor $\nu$.
(b) Incommensurability $1/(2d)$ of vertical stripes and uniform
spirals as a function of doping. Diamonds are data for LSCO from Ref.
\cite{yam98}; full dots are data for Fe-codoped LSCO from 
Ref. \cite{he10}. Results are obtained from the one-band model with 
parameters: $U/t=8$ and $t'/t=-0.2$.}
\label{bind}
\end{figure}

The task is now for a given parameter set and doping 
to variationally determine the lowest energy state among
the various stripe states and to compare with other possible solutions.
This has been accomplished in Refs. \cite{sei04,sei07,sei11} for the
one-band hamiltonian Eq. \ref{HM} and in Ref. \cite{lor02} also for 
the three-band model. Fig. \ref{bind} displays the binding energy
per hole
\begin{equation}
e_h = \frac{E_{AF}(N_h=0) - E_{text}(N_h)}{N_h}
\end{equation}
for the energetically most competing structures, namely stripes,
spirals and spin polarons. 
Here $E_{AF}(N_h=0)$ denotes the energy of the undoped AF and
$E_{text}(N_h)$ is the energy of a given texture obtained for doping
the system with $N_h$ holes.
We can also define the filling factor $\nu$ of stripes which corresponds
to the concentration of holes per unit cell along the stripe.
This can be related to the hole concentration $n_h$ and the distance
(in $a_{tet}$  lattice units)
between charge stripes $d$ as $\nu=n_h\cdot d$.
From the inset to Fig. \ref{bind}a it can be seen that for $U/t=8$ and
$t'/t=-0.2$ low doping
vertical stripes have $\nu_{opt} \approx 0.55$ whereas the minimum
of diagonal SC (BC) domain walls is at $\nu_{opt}=1 (0.75)$. We will
see below that  $\nu_{opt}$ determines the slope of the Yamada plot at low doping.

We find that for the one-band
model, vertical BC and SC stripes are also practically degenerate in energy 
in agreement with Refs. \cite{WS,fle01b,mar00}.
For definiteness we mainly restrict ourself to the BC case because
these textures constitute
the more stable configuration at $n_h=1/8$  in the more accurate
three-band model \cite{lor02} 
and in first principle computations\cite{anisimov04}.
However, one should keep in mind that all energetic considerations
below hold equally for SC vertical stripes.

From the inset to Fig. \ref{bind}a it turns out that diagonal 
BC stripes are practically 
(accidentally) degenerate in energy with vertical stripes. 
The energy of the DSC texture is
$\approx 0.02t$ per hole above. These small
energy differences should not be significant given the simplicity of the model.
A precise determination of the relative
stability of the different phases would require at least multiorbital
effects, inclusion of both long-range Coulomb interactions
and coupling of the holes to the tilts of the CuO$_4$ octahedra. The latter
have been shown to play a major role in the stabilization of
vertical {\it vs.} diagonal stripes\cite{normand01}.

In Fig.~\ref{bind} the dot-dashed line corresponds to the envelope
of the stripe binding energy curves for different periodicites $d$.
The solid (dashed) lines show $e_h$ for diagonal (vertical) spiral
solutions which apparently are unstable towards phase separation at low
doping. The horizontal line is the energy of the phase separated state
obtained from the Maxwell construction.
The respective stability of stripes and uniform spirals is determined
by the ratio between next-nearest neighbor and nearest neighbor hopping
$t'/t$ \cite{sei11,rac06,rac07}.  For $t'/t=-0.2$ stripes are 
stable over the whole doping range up to $n_h\approx 0.28$
where their energy becomes degenerate with that of vertical spirals.
In contrast, it is found that with increasing $|t'/t|$ stripes with
significant spin canting start to dominate the phase diagram and for
even larger $|t'/t|$ also checkerboard textures may become the ground state
\cite{sei07,sei11}. Also the optimum filling $\nu_{opt}$ changes with $t'/t$  
as expected according to the results of Ref.~\cite{sei04}.
The influence of a next-nearest neighbor hopping term on the
stripe stability has been previously investigated by
various methods. From the DMRG approach applied to the t-t'-J model \cite{WS3}
it turned out that a negative $t'/t$ can suppress both the
formation of stripes and pairing correlations.
A weakening of stripe tendencies for $t'/t<0$ in the same model
was also found with exact diagonalization
\cite{TOHYAMA} and in the Hubbard model with  DMFT \cite{fle01b},
and the HF approximation.~\cite{ichi99,val00,norm02}
All these calculations suggest that static stripes are destabilized
when the ratio $t'/t$ becomes negative.
This may indicate
a more dynamical character of stripes in some systems like
Tl and Hg based compounds.

With regard to the experimental evidence of stripe correlations in lanthanum
cuprates it turns out that a ratio of $t'/t=-0.2$ can account 
for the doping dependent incommensurability $\delta$ observed in
these compounds \cite{yam98}. For vertical stripes this quantity 
is related to the
modulation of the AF order and thus given by 
\begin{equation}\label{eq:epsvsdelta}
\delta=1/(2 d)=n_h/(2\nu)
\end{equation}
in units of $2\pi/a_{tet}$ where $a_{tet}$ is the tetragonal lattice constant.
The charged core of the stripe has a characteristic width $\xi$ so that
when the charge periodicity $d$ is larger than $\xi$ there are
negligible interstripe interactions and doping proceeds by increasing
the number of stripes.  In this regime the number of stripes for a
fixed number of holes is optimized when $\nu=\nu_{opt}$. Thus 
$\nu_{opt}$ determines the slope of the
incommensurability as a function of doping, according to
Eq.~(\ref{eq:epsvsdelta}) and $\nu_{opt}=0.55$
implies $\delta= 0.91 n_h$ in good agreement with the Yamada plot. 
The regime $d>\xi$ is characterized by the fact that  
$e_h$ at the minimum is independent of doping or $d$.
From Fig.~\ref{bind} we find $\xi=4$ lattice units which coincides
 with a direct examination of the charge profile. 
 For $d\lesssim
 \xi$ stripes completely cover the plane and increasing further the number of
 stripes becomes energetically costly. Therefore doping proceeds by
 increasing the charge of stripes and $\delta$ becomes constant. 
The crossover in doping between the two regimes occurs when 
$n_h=\nu_{opt}/d  \sim
1/8$ in good agreement with the Yamada plot in LCO. 

As can be seen from Fig. \ref{bind}b the incommensurability for stripes 
shows a staircase structure where the steps occur
at the crossing of the  corresponding energy curves
(cf. Fig. \ref{bind}a). For small doping ($d>\xi$), 
since interstripe interactions
are negligible, one can produce a practically continuous curve by
considering combinations of solutions with
periodicity $d$ and $d+1$. As explained above 
this is not any more convenient 
for $d\lesssim \xi$ and 
the additional holes are doped into existing stripes producing the
plateau in the incommensurability. 

Contrary to the stripes, one             
observes a continuous increase of $\delta(n_h)$ in case of uniform spirals .  
Nevertheless, both phases show similar evolution of incommensurability 
with doping and indeed fairly similar values of $\delta(n_h)$ 
up to $n_h \sim 1/8$. For larger doping this agreement is 
lost due to the tendency of stripes to make wide plateaus.

We have explained the Yamada plot in terms of well formed stripe
textures. It is remarkable that the incommensurabilities, both for
stripes and spirals, are not far
from what one would obtain from an analysis of the momentum dependent 
susceptibility of the Fermi liquid phase \cite{comnest}.
In fact, if one adiabatically decreases the interaction  
both uniform spirals and stripes originate from
the same magnetic instability\cite{marki10}. One can call this a ``weak
coupling'' instability. On the other hand
within the
time-dependent Gutzwiller approximation, the Hubbard interaction $U$
is strongly screened by vertex corrections. Therefore a quite large
value of $U$ is needed in order to make the Fermi liquid unstable and
the instability is better characterized as ``intermediate coupling''. 

The time dependent Gutzwiller analysis shows that it is 
the plateau in the spin susceptibility close to $Q_{AF}=(\pi,\pi)$
(cf. Fig. 1 in Ref. \cite{marki10}) which drives the system unstable
towards spiral or stripe order for sufficiently large on-site repulsion
$U/t$. 

Figure~\ref{bind}b also shows data from more recent neutron scattering
experiments \cite{he10} on Fe-LSCO. In these overdoped samples an elastic
incommensurate spin response was found close to the dominant nesting vectors
as extracted from ARPES experiments. It was thus concluded that the induced
incommensurate response signals an inherent instability of the itinerant
charge carriers, being different from the low doping stripes arising
from localized Cu spins.
However, from Fig. \ref{bind}b it turns out that the data are rather close
to the incommensurability curve of stripes obtained for the
LSCO parameter set. Remarkably this behavior was predicted in
Ref.~\cite{lor02} much before the measurement became available.   
Since these stripe calculations are based on an
itinerant approach, there is no `dichotomy' between low doping `localized'
spin stripes and large doping itinerant ones, but both appear as different
limits of the same model.

Finally we briefly address the incommensurability of diagonal stripes.
In tetragonal units the incommensurability is a factor of
$\sqrt{2}$ larger than that for vertical stripes given
in Eq.~(\ref{eq:epsvsdelta}). The more stable BC diagonal textures
have $\nu_{opt}\approx0.75$ so that $\delta_{diag}=n_h/(\sqrt{2}\nu_{opt})
\approx n_h/1.06$. Therefore the doping dependence of $\delta$ for BC diagonal
and vertical stripes is almost identical in agreement with neutron scattering data \cite{waki99,waki00,mats00,fujita02}.
One should keep in mind the effect of disorder which in the
diagonal phase reduces the correlation to  the same order or even smaller 
than the periodicity\cite{waki00} whereas for vertical stripes the
correlation length can reach $150 a_{tet}$ or around 20 times the
magnetic stripe periodicity.  This becomes especially important in
context of spin excitations from stripes in the spin glass phase
(cf. Sec. \ref{sec4}).

\section{Electronic structure and transport properties}\label{sec3}
\begin{figure}[htb]
\includegraphics[width=7.5cm,clip=true]{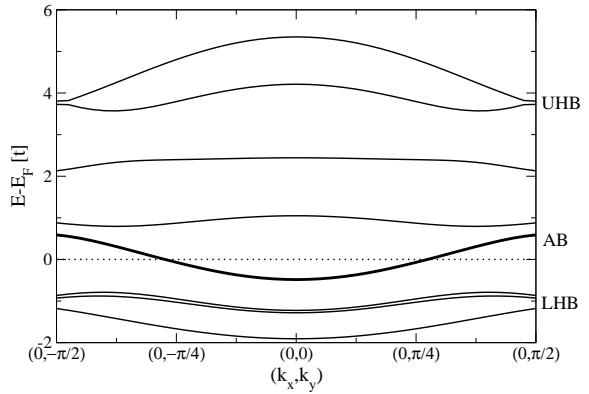}
\caption{Band structure for $d=4$ bond-centered stripes
oriented along the y-direction. The dispersion is shown
for a cut along the stripe and $k_x=0$.
The electronically relevant active band (AB) crosses the Fermi energy
(doping $n_h=0.125$) and is in the gap between lower (LHB) 
and upper (UHB) Hubbard band. The AF modulation in the direction of
the stripes leads  to a doubling of the periodicity so that 
$-\pi/2 \le k_y < \pi/2$.}
\label{fig:bands}
\end{figure}

Fig. \ref{fig:bands} shows the bandstructure for half-filled
$d=4$ bond-centered stripes which are oriented along
the y-direction. The electronic states of the undoped AF ordered
regions (magnetization per spin $m$) give rise to lower and upper 
Hubbard bands (LHB/UHB) which are separated by $\Delta\sim U\cdot m$.
Stripes induce the formation of additional bands in the gap and
for partially filled stripes (i.e. $0<\nu<1$) one of them
('active band')  crosses the Fermi energy.
For the present case of 'half-filled' stripes the crossing occurs
at $k^F_y=\pm \pi/4$ for $k_x=0$ ($a_{tet}=1$). Perpendicular to the
domain walls the bands are rather flat with the residual small dispersion
due to the overlap between adjacent stripes.

\begin{figure}[htb]
\includegraphics[width=7.5cm,clip=true]{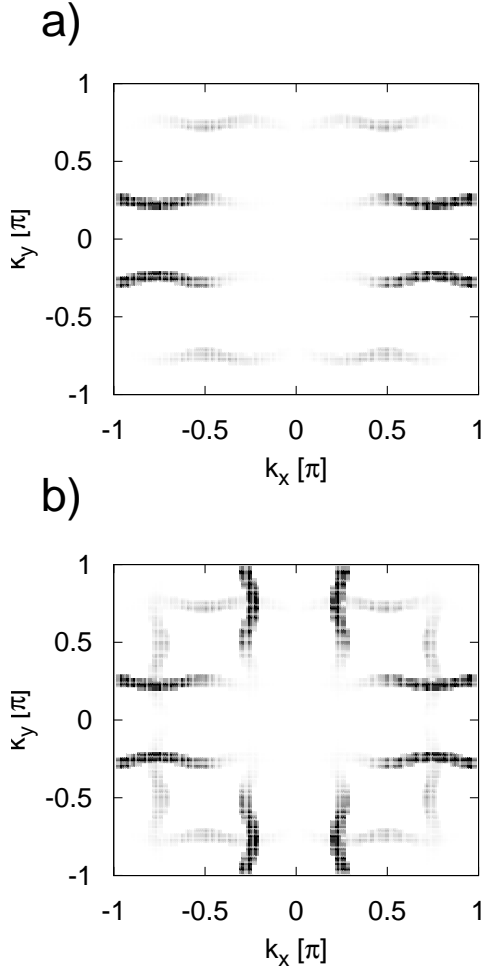}
\caption{Fermi surface for $d=4$ bond centered stripes at $n_h=0.125$.
(a) y-axis oriented stripes; (b) superposition of x- and y-axis oriented
stripes.}
\label{fig:fs}
\end{figure}

The 'active band' implies a quasi one-dimensional Fermi surface 
at momentum $k_y=\pi/4$ (cf. Fig. \ref{fig:fs}a). The corresponding
weight in the full Brillouin zone is concentrated around the
$(\pi,0)$ point which results in the two parallel FS segments.
There is also a crossing at $k_y=3 \pi/4$ with less spectral weight
and the associated segment is shifted more towards $k_x=0$.
In LSCO one expects a superposition from x- and y-oriented stripes
which results in the FS shown in Fig. \ref{fig:fs}b. Such a
prediction is in good agreement with ARPES data \cite{zho99,zho01}
and even fine details as the small curvature of the segments due
to interstripe hopping are reproduced.
It should be mentioned that the experimental data \cite{zho99,zho01}
strongly depend on the energy window which is chosen to extract the
FS. We will address this point in Sec. \ref{sec5}

As discussed above, stripes are approximately half-filled for doping
$n_h \lesssim 1/8$ and for $n_h \gtrsim 1/8$ additional holes are
doped into the active band. Therefore the chemical potential
is expected to be approximately  
constant for $n_h \lesssim 1/8$ and decreases for $n_h \gtrsim 1/8$ in     
qualitative agreement with the observed behavior\cite{ino97,har01}.    
The rate of change of $\mu$ with doping, being a high derivative of    
the energy, is very sensitive to finite size effects and, moreover,  few
experimental points are available in this doping range in order to allow for
a precise comparison. A rough estimate indicates that the theoretical       
rate of change of $\mu$ with doping for $n_h>1/8$ is approximately a factor   
of 2 larger than the experimental one\cite{ino97,har01}. This may be        
attributed to an underestimation of the mass renormalization in             
mean-field. Another possibility                                             
which goes in the right direction is phase separation                       
among the $d=4$ stripe solution  and the paramagnetic overdoped Fermi       
liquid which is also suggested by the doping evolution of magnetic excitations
(cf. Sec. \ref{sec4}).                                                    

We finally note that also anomalies in the Hall and Nernst coefficient
can be attributed to the FS reconstruction due to striped
ground states \cite{lor02,martin10,hackl10}.

\begin{figure}[htp]
\includegraphics[width=5cm,clip=true]{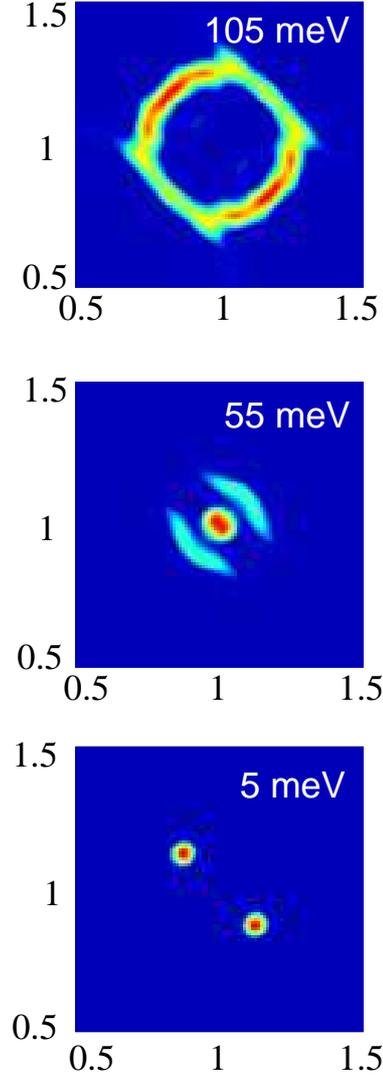}
\caption{Constant frequency cuts of the magnetic excitation spectrum at
$\omega=5, 55$, and $105 meV$ for the $d=4, n_h=0.125$ BC stripe structure.
The momentum space corresponds to the magnetic Brillouin zone which
is rotated by $45^0$ with respect to the stripe direction (cf. Fig. \ref{fig:bz}).}
\label{fig:int}
\end{figure}

\section{Spin excitations}\label{sec4}
\begin{figure}[htb]
\includegraphics[width=7cm,clip=true]{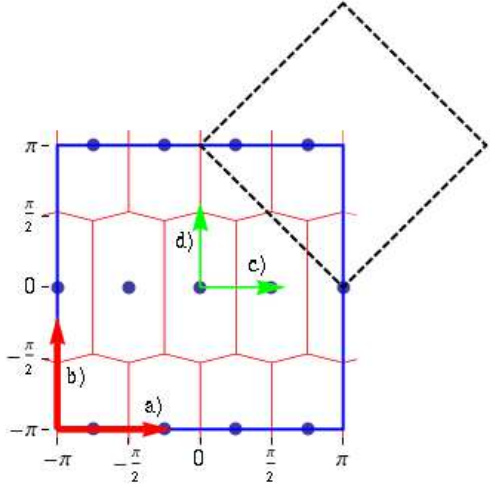}
\caption{Subdivision of the full Brillouin zone into reduced zones
defined from the elementary cell of y-axis oriented $d=4$  BC stripes.
Arrows (a) ... (d) indicate the scan direction for
the magnetic dispersions shown in Figs. \ref{fig:disp}  and \ref{fig:rixs}.
The intensity plots of Fig. \ref{fig:int} are shown within the
magnetic Brillouin zone indicated by the dashed square.}
\label{fig:bz}
\end{figure}

Fig. \ref{fig:int} reports constant frequency scans of the imaginary part
of the transverse magnetic susceptibility
$\chi^{''}_q(\omega)=-Im \frac{1}{N} \int e^{i\omega t}
\langle {\cal T} S^+_q(t) S^+_{-q}(0)\rangle$ 
for y-axis oriented $d=4$ BC stripes in the magnetic Brillouin zone
(cf. Fig. \ref{fig:bz}).

\begin{figure*}[htb]
\includegraphics[width=16cm,clip=true]{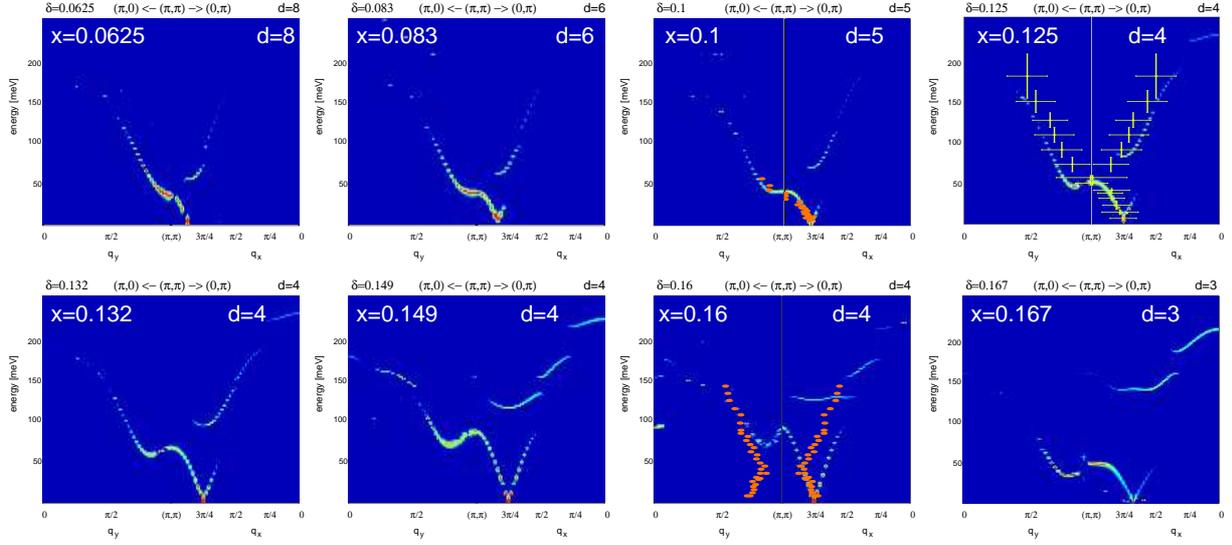}
\caption{Dispersion of spin excitations perpendicular
and along y-axis oriented BC stripes for various dopings. Experimental data
are from Ref. \cite{kofu07} ($n_h=0.1$) , Ref. \cite{tra04} 
($n_h=0.125$), and Ref. \cite{vignolle} ($n_h=0.16$).}
\label{fig:disp}
\end{figure*}

The low frequency intensity is concentrated around the 
wave vectors $Q_s=(\pi(1\pm 1/4,\pi)$                                    
corresponding to the spin periodicity of the underlying                  
$d=4$ stripe structure. In fact, it turns out that
the weight of the elastic magnetic peaks is much
larger at $Q_s$ than at the higher harmonics (cf.
Fig. \ref{bragg} where the corresponding weights obtained from the 3-band
model calculations are shown). Note also that the magnetic weight vanishes
identically on the 'nuclear' Bragg points $Q_n= n (\pi/2,0)$
which only contribute in the charge channel.
As a result the dispersion along the cut $(q_x,\pi)$ 
(cf. Fig. \ref{fig:disp}, $n_h=0.125$)              
shows the expected behavior for spin-waves in a striped system           
\cite{sei06,kru03,car04,car06}. Starting from the Goldstone mode at $Q_s$ 
one observes two branches of spin waves where the large intensity one    
disperses towards $Q_{AF}=(\pi,\pi)$. The other branch rapidly looses intensity
which is analogous to what is found within linear spin-wave                    
theory of the Heisenberg model \cite{sei06,kru03,car04,car06}
for small ratios of $J_\perp/J_\parallel$.                                     
Here $J_\perp$ denotes the (ferromagnetic) exchange coupling across a stripe   
whereas  $J_\parallel$ denotes the coupling within the AF ordered legs.        
In addition the intensity of the outwards dispersing branch is 
strongly reduced by disorder \cite{boothroyd11}.

At $E\approx 55meV$ the spin excitations have finally shifted
towards the AF wave-vector producing an intense feature at $Q_{AF}$
(middle panel of Fig. \ref{fig:int}) which is associated with a 
saddle-point in the magnetic dispersion [Fig. \ref{fig:disp} ($n_h=0.125$)].
At the same time one observes 
the appearance of intensity in the stripe direction starting from $Q_{AF}$.
From Fig. \ref{fig:disp} ($n_h=0.125$) it turns out that this
feature is due to a slight softening of the magnetic excitations
upon dispersing away from the saddle-point in the direction of the stripe.
This softening is absent at lower doping [cf. Fig. \ref{fig:disp} ($n_h=0.0625$,
$n_h=0.083$)] and indicates the susceptibility towards a magnetic instability
along the stripe at higher doping. It has been argued \cite{sei06} that 
this roton
like minima explains INS in  untwinned samples of YBCO by Hinkov and
co-workers \cite{hin04}.

Finally, at higher energies the excitations again spread out and           
form a ring-shaped feature around  $Q_{AF}$ with a small anisotropic         
intensity modulation (Fig. \ref{fig:int}, upper panel). 
Note that our spectra are                              
for one-dimensional stripes without any orientational average                
so that C$_4$ symmetry is broken.
The resulting quasi two-dimensional excitation spectra 
above the saddle-point energy
are in marked contrast to the one-dimensional characteristics
obtained from localized spin models \cite{voj04,uhr04}.
This has profound implications on the interpretation of
INS data from detwinned samples \cite{hinkov07,hinkov08} which also show
pronounced 1-D characteristics at low but a 2-D magnetic distribution
of magnetic energies at high energies in accord with our result.

The high energy distribution of magnetic intensity 
can be understood from the energy weighted sum rule                   
\begin{equation}                                                    
M_{\bf q}^1\equiv\int_0^\infty \! d\omega \,\omega \chi_{\bf q}''(\omega) 
= -\frac{\pi}{2 N}\sum_{{\bf k},\sigma}                                   
(\varepsilon_{\bf k}- \varepsilon_{\bf k+q}) n_{{\bf k},\sigma}           
\label{eq:sr}                                                             
\end{equation}                                                            
where $\varepsilon_{\bf k}$ denotes the single-particle dispersion
of the underlying itinerant model and $n_{{\bf k},\sigma}$
is the number operator for particles with momentum ${\bf k}$ and spin $\sigma$.
In the presence of time reversal symmetry and inversion                                  
($n_{{\bf k},\sigma}=n_{{\bf -k},-\sigma}$) this can be                   
simplified to                                                             
\begin{equation}                                                          
M_{\bf q}^1 = \frac{\pi}{N}\sum_{\delta}\sin^2\left(\frac{{\bf q}\delta}  
{2}\right)T_\delta .                                                      
\label{eq:sr2}                                                            
\end{equation}                                                            
and $\delta$ runs over the vectors connecting a specific lattice 
site to its respective neighbors with only one representative of the 
pair $(\delta,-\delta)$.     
Thus in case of a square lattice $\delta = \hat{\bf x},\hat{\bf y}$        
for nearest neighbors,                                                     
$\delta = \hat{\bf x}+\hat{\bf y}, \hat{\bf x}-\hat{\bf y}$ for            
next nearest neighbors etc.                                                
and $T_\delta$ denotes the kinetic energy in $\delta$-direction.           
Since $M_{\bf q}^1$ weights more the excitations at higher energies        
it turns out from Eq. (\ref{eq:sr2}) that with increasing $\omega$ the      
intensity in the transverse magnetic susceptiblity should be confined      
to wave-vectors around $Q_{AF}=(\pi,\pi)$. 

Consider now the case of stripes oriented along the y-direction.
Although we are dealing with a perfectly ordered quasi 1D structure
we find from our calculations that e.g. for $d=4$ BC stripes the   
kinetic energy along the y-direction is only $\sim 10\%$ larger than
that perpendicular to the stripes.
Neglecting the contribution from next-nearest neighbor terms
Eq. (\ref{eq:sr2}) becomes
\begin{equation}
M_{\bf q}^1 = \frac{\pi T_{\hat{\bf x}}}{N}\sin^2\left(\frac{{q_x}\delta}
{2}\right)+ \frac{\pi T_{\hat{\bf y}}}{N}\sin^2\left(\frac{{q_y}\delta}
{2}\right)
\end{equation}
so that even in the presence of stripes the high energy magnetic
response is essentially two-dimensional but slightly anisotropic
around $Q_{AF}$.
In Fig. \ref{fig:int} this is exactly the distribution of the
optical magnetic excitations one finds above the saddle-point energy.
We anticipate that this
feature can also be observed in the spin glass-phase of LSCO where due
to the imbalance of twin domains the one-dimensional character
of low energy spin excitations has been resolved by inelastic neutron
scattering experiments \cite{matsuda11}. Above $\omega_{s}$ the
measurements seem to indicate that the excitations acquire
again the (two-dimensional) character of the magnons in the
undoped system in agreement with the prediction for a striped
ground state \cite{sei09}.

Another interesting aspect of the sum rule
Eq.~\ref{eq:sr}, that has 
 passed unnoticed so far, is that it allows to measure the change
in kinetic energy on entering the superconducting state in a way that
is alternative to the proposal in Ref.~\cite{sca98}. The present
proposal needs accurate measurements in the higher part of the spectra
which should become available with the new pulsed neutron sources.

The overall doping evolution of the vertical stripe magnetic excitations
together with availabe experimental data is depicted in Fig. \ref{fig:disp}.
Here we show the excitations along scans (a,b) sketched in Fig. \ref{fig:bz}.
It should be noted that the magnetic excitations for SC stripes
are almost identical
with only slight differences regarding the intensity distribution
and the gap structure of optical branches \cite{sei05}.
It turns out from Fig. \ref{fig:disp} that for doping $n_h>0.125$ the
saddle-point excitation $\omega_{s}$ at $Q_{AF}$ rapidly 
shifts to higher energies together with an
increasing softening of the magnetic excitations along the stripe.
In fact, the value of $\omega_s$ is determined by the magnetic coupling
across the domain wall. Because this coupling is mediated by virtual
hopping processes of holes between stripe sites and adjacent AF regions
it strongly depends on the stripe filling factor $\nu$ as defined
above. As we have shown above for $n_h<1/8$ the filling of the core of
the stripe remains practically constant and $\omega_s$ has a weak
dependence on doping. On the other hand 
 for $n_h>1/8$ doping proceeds by changing the filling of the stripes  
at constant incommensurability. This produces a rapid renormalization
of the effective exchange interaction across the core and a
concomitant rapid increase in $\omega_s$.

\begin{figure}[htb]
\includegraphics[width=7cm,clip=true]{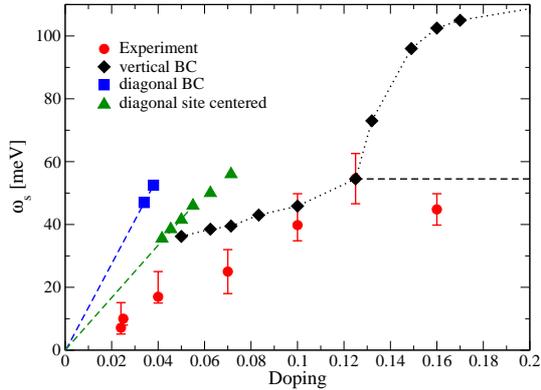}
\caption{Doping dependence of the saddle-point excitation
$\omega_{s}$ at $Q_{AF}$ for diagonal site- and bond-centered stripes
and vertical bond-centered stripes.  Circles with error bars are               
experimental data from  
Refs. \cite{tra04,hiraka01,vignolle,kofu07,matsuda08,matsuda11}. 
The horizontal dashed line starting from $n_h=0.125$ is the expected
behavior for a phase separation scenario for stripes.}
\label{fig:wres}
\end{figure}

In Fig. \ref{fig:wres} we compare the doping dependence of $\omega_{s}$
with available experimental data
\cite{tra04,hiraka01,vignolle,kofu07,matsuda08,matsuda11}. The latter
seem to indicate \cite{vignolle} that in LSCO 
the saddle-point excitation at $Q_{AF}$  stays at least constant
beyond $n_h=1/8$ 
in contrast to what is obtained from the spectra of doped stripes shown
in Fig. \ref{fig:disp}.
It is conceivable that instead of doping additional holes
into the domain walls these (fluctuating) textures in LSCO stay
half-filled  beyond $n_h=1/8$ leading to phase separation between
free carriers and those which are involved in the stripe correlations
\cite{lor02}.  On the other hand Wakimoto and
collaborators\cite{wak07} find that at large doping $n_h=0.3$
scattering starts around  80meV which is compatible with a strong
increase of  $\omega_s$ as we predict for the doped stripes. Thus more
experimental work is needed at intermediate dopings to clarify the situation.

With underdoping the computations increasingly overestimate the
energy of the experimentally determined $\omega_s$, especially in the 
diagonal phase. The main reason for the disagreement in this doping range is 
probably the increasing disorder character of the stripes.
To support this idea we have performed \cite{sei09} linear spin wave 
theory calculation of disordered diagonal bond centered stripes.
Is is found that already a small amount of disorder leads to a 
softening and broadening  of $\omega_{s}$  which would bring the
theoretical data in Fig. \ref{fig:wres} closer to the experimental ones.

\begin{figure}[htb]
\includegraphics[width=7.5cm,clip=true]{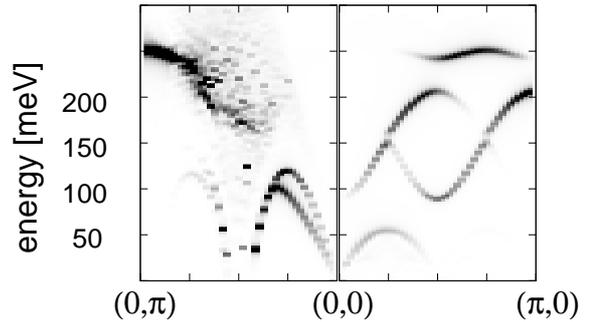}
\caption{Magnetic excitations in the nuclear Brillouin zone 
for $d=4$ y-oriented stripes. The spectra correspond to the scans
labeled (c,d) in Fig. \ref{fig:bz}.}
\label{fig:rixs}
\end{figure}

Recent experimental effort has been made to measure magnetic
excitations in LSCO with resonant inelastic x-ray scattering \cite{brai10}.
Because this technique essentially measures in the nuclear Brillouin
zone (i.e. close to momentum ${\bf q}=0$) we show in Fig. \ref{fig:rixs}
the corresponding spectra for $d=4$ y-axis oriented bond centered stripes.
Consider first the scan starting from momentum ${\bf q}=(0,0)$ 
[labeled (d) in Fig. \ref{fig:bz}] in the
direction along the stripe. Because the system is still AF ordered along
the stripe (cf. Fig. \ref{ststruc}) one has a doubling of the unit
cell in this direction. At the same time the doping of the stripe 
is approximately $\nu \approx 0.5$ so that the low energy structure
of the spectra reflects the continuum of spinon like spin-flip
particle-hole excitations for 
the active band shown in Fig. \ref{fig:bands}.  
At higher energy, part of the (high intensity) optical magnon band 
overlaps with the spin-flip continuum and dominates the spectrum.
In the direction perpendicular to the stripe
scan (c) is connected to scan (a) by the addition of a reciprocal lattice vector
$(3\pi/4,\pi)$. Therefore the low energy magnon band which disperses
from $(0,0)$ to ($\pi,0)$ is the replica of the  Goldstone mode which
has been discussed above. The same holds for the higher energy branches
which, however,  in the nuclear and magnetic zones strongly differ
in weight.

\section{Optical conductivity}\label{sec5}
In the previous sections our considerations where mostly based on the
extended one-band model, which should be appropriate as long as
interband transitions (as in the case of spin excitations) do not
play a significant role. In the present section we discuss the
optical conductivity of striped systems which involves also higher
energy excitations. We therefore investigate the corresponding spectra
in the framework of the three-band model
which is again treated within the time-dependent Gutzwiller approach
\cite{lor02,lor03}. 

It is also possible to obtain an optical
conductivity that in the main features resembles experiment, using a
one-band Hubbard model. Indeed, one could certainly argue from the
inset to Fig. \ref{fig:edq}b  that the 
transition between LHB and UHB  mimics the 
charge-transfer excitation of cuprates and therefore should provide
a reasonable starting point. However, the point is that the charge excitations
will also be strongly influenced by the respective stabilitity of
bond- and site-centered stripes which are energetically degenerate
in the one-band model. This leads to soft lateral fluctuations of the
stripes (phason mode) which are optically active. Because of the
degeneracy between the two textures, the phason is soft in the one-band
model at low doping. In contrast, in the more realistic three-band model 
the two textures become only degenerate at higher doping  
which, as shown below, is in good agreement with experiment.

\begin{figure}[htb]
\centering
\includegraphics[width=7cm,clip=true]{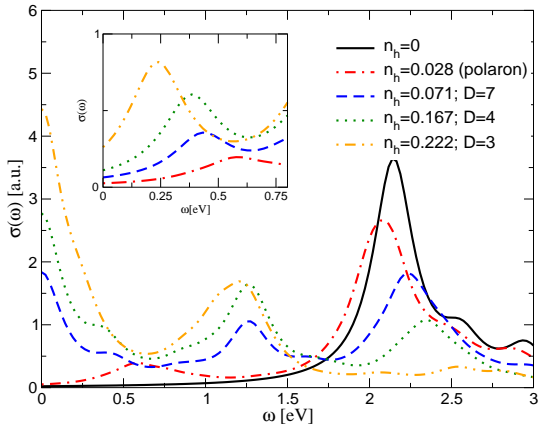}
\caption{Optical conductivity of the three-band Hubbard model
evaluated within the time-dependent GA for various hole concentrations.
In the inset: The evolution of the mid-infrared peak as a function of doping.
The Drude contribution has set to zero in this case.
}
\label{optcond}
\end{figure}

Based on the three-band hamiltonian Fig.~\ref{optcond} reports the optical 
conductivity for AF and inhomogeneous ground states at various
dopings.         The parameter set has been taken from the first
principle computations of Ref. \cite{mcm90} so we have no adjustable
parameters. 
     
At very dilute doping ($n_h \lesssim 0.03$)                        
due to the long-range Coulomb interaction (not included in our   
calculations) each hole will be close to an acceptor preventing  
the formation of stripes.                                        
The RPA optical conductivity for the corresponding single-hole solution 
($n_h=0.028$) leads to the formation of a                            
doping induced MIR band close to 0.5eV and doping induced transfer      
of spectral weight from the charge transfer (CT) band to the MIR region 
in agreement with experiment in this doping range\cite{uch91}.          

For distant stripes ($d=7$) the single-hole MIR band now splits
into two bands. The one at higher energy is a band of incoherent
particle-hole excitations close to 1.3eV which provides a       
theoretical explanation for the shoulder to the CT excitation   
seen at the same                                                
energy in optical absorption through LSCO thin films \cite{suz89},
and electron energy loss spectroscopy \cite{ter99}.               

The low energy MIR peak is a collective mode associated with the soft
lateral displacements of the stripe mentioned above. 
It is located
at 0.3 eV at low doing and becomes soft only at optimum doping
where it  merges with the Drude
response in good agreement with experiment. This soft collective modes are
low energy excitations which have no counterpart in a noninteracting
system breaking the paradigm of Fermi liquid theory \cite{lor03}.  It is also
possible that coupling to these modes contributes significantly to the
pairing. It is worth to remark that new developments in time resolved
spectroscopy show \cite{man11} that the high energy excitations below 
the CT band
do not participate in the pairing but excitations at the CT band do.
The region below 1.6eV has not been measured yet. 

It is interesting to remark that also in nickelates the MIR peak has
been attributed to the formation of mid-gap states due to the stripe
ground state \cite{homes03}.

\section{Dynamical stripes}\label{sec6}
The investigation of stripe 'dynamical properties' in the previous
sections concerned the excitations on top of a stripe
ground state with long range order. 
However, as discussed in Secs.~\ref{exp},\ref{sec4} in most cuprate materials
long range static spin correlations are not observed and 
also direct evidence for static charge order is quite elusive.

Local probes like STM often point to a broken translational symmetry
in real space specially when the high energy part of the spectra is
observed. Momentum space probes like  ARPES  and INS
 resemble those of an ordered state at high energy but interpolate to
 a  state without broken symmetry at low energy.  This 
 points towards a dynamical or a glassy nature of the stripe
correlations. 

There are two ways in which this dichotomy between low and high energy
quasiparticle excitations can be solved: 1) proximity to a
quantum/classical 
critical point in the disordered side (Sec.~\ref{sec:prox-quant-crit}). 2) Stripes with long range
order but with protected low energy quasiparticles (Sec.~\ref{sec:protected-low-energy}).  For the last
possibility there are also two variants: 2.a) Long range order in spin
and charge and 2.b) Long range order only in the charge.

\subsection{ Proximity to a quantum/classical critical point}
\label{sec:prox-quant-crit}

The classical version of scenario 1) consists of thermally disordered
stripes at finite temperature. This has been the route followed by 
Vojta and collaborators \cite{vojta06} who have investigated a 
Landau model for coupled charge and spin fluctuations in the
Born-Oppenheimer approximation. Although not mentioned explicitly, this
computation also describes a glassy stripe state.

The quantum version of scenario 1)  follows from this qualitative
argument: In physical                                             
 dimensions a system may have long (but finite) ranged order      
parameter spacial correlations which are also long lived close to a 
quantum critical point. This defines a fluctuating frequency        
$\omega_0$ above which the systems appears to be ordered (or at least
critical).  Then,
for energies larger than $\omega_0$ with respect to the Fermi  
level, the spectral function should resemble the spectral function of an       
ordered system. This high energy spectral weight which normally one
would term ``incoherent'' may in reality  
 carry important information on the momentum  
structure of the close-by ordered phase. This momentum structure may be
determined with ARPES from the momentum distribution  
$n_{\bf k}$ by integrating the spectral function over a broad energy window
or by STM and INS experiments by direct examination of the high energy
spectra.

On the other hand, electrons at lower energies average                  
over the order parameter fluctuations and ``sense'' a disordered       
system. In this limit we expect Fermi liquid quasiparticles            
with all their well known characteristics like a Luttinger Fermi       
surface. This dichotomy in the momentum structure of low and high 
energy quasiparticles is consistent with 
ARPES data \cite{zho99,zho01} on lanthanum cuprates.

\begin{figure}[htb]
\includegraphics[width=7cm,clip=true]{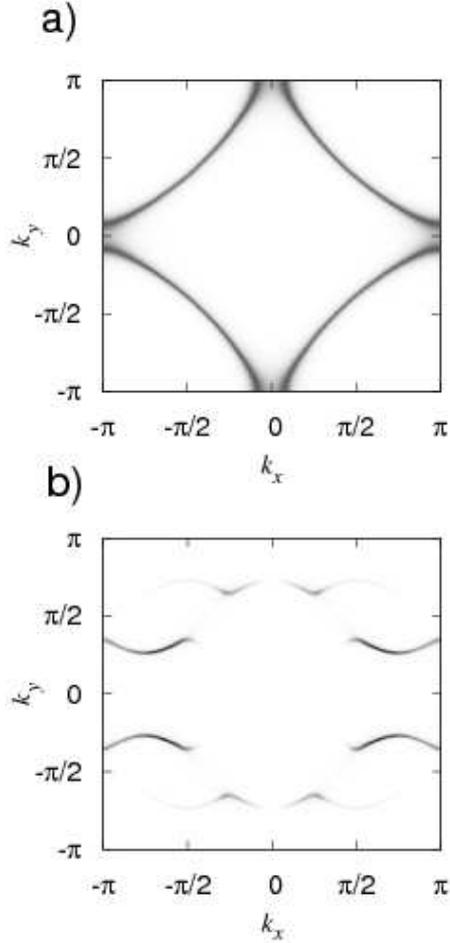}
\caption{Constant frequency cuts of the electronic dispersion for dynamical
$d=4$ bond centered Gutzwiller stripes.
(a) $\omega = E_F$, (b) $\omega=-0.6 t$. Parameters: $U/t=8$, $t'/t=-0.2$.
The frequency structure is obtained from Eq. \ref{eqs3} with $\omega_0=0.2t$
and $v^2=0.02$.}
\label{fig:dyn}
\end{figure}

In a series of papers \cite{sei01b,grilli05,grilli09,sei10} 
which were based on a Kampf-Schrieffer type approach \cite{KAMPF} 
we have worked 
out the scenario of fermionic quasiparticles coupled to 
dynamical charge order fluctuations
with regard to several experimental observations.
These include the angular dependence
of the quasiparticle weight in Bi2201 \cite{kondo09},
the isotope shift of the high-energy electronic dispersion \cite{GWEON},
and the dichotomy in the Fermi            
surface of high-T$_c$ cuprates \cite{zho99,zho01} as mentioned before.

The Kampf-Schrieffer approach describes quite successfully situations
in which the quasiparticles are coupled to locally well formed order-parameter
fluctuations but can not be derived from a microscopic approach. On
the other hand a formal derivation close to a quantum critical point
leads to  quasiparticles
coupled to diffusive collective modes (CM)s. This leads to a form 
which is customary in quantum critical phenomena 
and has often been used for spin fluctuations \cite{chubukov}
\begin{equation}
D_\lambda({\bf q}, \omega)=-\frac{1}{m_\lambda+\nu_\lambda
({\bf q}-{\bf q}_{\lambda})^2-i\omega-
\omega^2/{\overline{\Omega}_\lambda}} 
\end{equation}
where $\lambda=c,s$ refers to charge or spin CMs
substantially peaked at characteristic wavevectors ${\bf q}_s\approx (\pi,\pi)$ and
${\bf q}_c\approx (\pm \pi/2,0),(0,\pm \pi/2)$ respectively.
Here the mass $m_\lambda$ is the minimum energy required to excite the CM and
$\nu_\lambda$ is a fermion scale setting the CM momentum dispersion. This dispersion is
limited by an energy cutoff $\Lambda_\lambda$. The dimensionless quantities $m_\lambda/\Lambda_\lambda$ are
the inverse square correlation lengths (in units of the lattice spacing), which measure
the typical size of ordered domains.
The $i\omega$ term establishes the low-energy diffusive character of
these fluctuations due to decay into particle-hole pairs, whereas above
the scale set by $\overline{\Omega}_\lambda$ the CM has a more propagating character.
This approach has been used to investigate the critical behavior of the low-frequency 
optical conductivity \cite{ENSS} and, more recently,  to calculate the Raman response function 
including self-energy and vertex corrections\cite{CAPRARA}. In particular, 
assuming as mediators spin and charge fluctuations with different characteristic 
wavevectors, it was possible to selectively individuate the charge and spin  contributions 
in the Raman response function.
By fitting the experimental spectra in ${\rm La_{2-x}Sr_xCuO_4}$ single crystals within the above
theoretical framework, it was found that both charge and spin  fluctuations
were contributing to the quasiparticle scattering (a distinctive feature of stripe
fluctuations). However,  it was also found that, upon increassing doping, charge CMs
progressively acquire more relevance while spin fluctuations become weaker and weaker.
This indicates that upon increasing doping into the overdoped regime,
stripes naturally evolve into harmonic damped charge-density-wave fluctuations,
 while spin modes eventually loose weight.

\subsection{ Protected low energy quasiparticles}
\label{sec:protected-low-energy}

STM \cite{kohsaka08,alldredge08} experiments provide strong evidence
for the simultaneous existence of low energy (Bogoljubov) quasiparticles
and manifestations of truly broken translational symmetry at 
high energy.

In another  recent approach \cite{sei092,sei10} we have developed
a phenomenological description with broken translational symmetry but
a dynamical order parameter which applies to this situation.  
The theory also accounts for the contrast reversal in the STM spectra 
between positive and negative bias \cite{ma08}, in contrast to mean
field like static order where this effect in general occurs away from the
Fermi level.  
Here we show that this theory can be combined with the GA computations 
discussed in Sec. \ref{sec2}  in order to phenomenologically
account for the electronic structure of a fluctuating stripe state.

The GA is a renormalized mean-field theory which allows for the definition
of an effective Gutzwiller hamiltonian (cf. e.g.
Ref. \cite{goe03})
\begin{equation}\label{hga}
H^{GA}=\hat{T}+\hat{V}\equiv 
\sum_{ij\sigma}\widetilde{t_{ij}}c_{i\sigma}^\dagger c_{j\sigma}
+\sum_{i\sigma}\lambda_{i\sigma}c_{i\sigma}^\dagger c_{i\sigma}
\end{equation}
where the $c_{i\sigma}^{(\dagger)}$ are now (creation) annihilation operator
of the Gutzwiller quasiparticles. The $\widetilde{t_{ij}}$ are renormalized
hopping amplitudes and $\lambda_{i\sigma}$ correspond to local chemical
potentials. Both parameter sets depend on the value of the onsite repulsion but
also on the charge and spin structure of the inhomogeneous ground state.

We want to construct an ansatz for the self energy due to
electron-electron interactions. In order to ensure that the
self-energy is physical (i.e. that obeys all analytical properties of
an electronic self energy) we map the problem to that of electrons
coupled to a set of auxiliary states.
The idea is  to enlarge the Hilbert space by introducing the
coupling to an additional set of localized states (described by creation
and destruction operators $f_{n\sigma}^{(\dagger)}$ which thus results
in the following (auxiliary)  Fano-Anderson Hamiltonian,

\begin{displaymath}
H^{aux}=\hat{T} + \sum_{i,n,\sigma}v_{i,n} \left\lbrack c_{i\sigma}^\dagger f_{in,\sigma}
+ h.c. \right\rbrack + \sum_{i,n,\sigma}\varepsilon_{in}f^\dagger_{in,\sigma}
f_{in,\sigma}.
\end{displaymath}

As a result the Greens function for the Gutzwiller quasiparticles depends
on an additional self-energy which is determined by the electronic
structure of the $f$-states via 

\begin{equation}\label{eq:self}                                       
\Sigma_i(\omega)=\sum_n\frac{v_{i,n}^2}{\omega-\epsilon^f_{in}}+\Delta_i
\equiv f_i(\omega)+\Delta_i.                                 
\end{equation}                                                        
with the constant $\Delta_i$ controlling the high frequency limit.

Note that this kind of approach automatically maintains the
correct analytical properties of the (frequency dependent) self-energy 
which replaces the static single-particle potential in Eq. \ref{hga}.

For the frequency structure in Eq. \ref{eq:self} we choose
a two-pole Ansatz

\begin{eqnarray}                                                  
f_i(\omega)&=&\left\lbrack \frac{\alpha_i}{\omega-\omega_0}       
+\frac{\beta_i}{\omega+\omega_0}\right\rbrack  \\                 
\Delta_i&=& \frac{\alpha_i-\beta_i}{\omega_0} \label{eqs3}        
\end{eqnarray}
with $\alpha_i+\beta_i=v^2$ and $v^2$ controls the strength of the scattering.
Thus at each site an asymmetry is introduced in
the spectral distribution which below is taken to be proportional to the
charge modulation. The constant $\Delta_i$ guarantees the limit
$\Sigma_i(\omega=0) = 0$ at each site.
Denoting by $\overline{\lambda}$ the average local
chemical potential the weights of the poles are choose as
$\alpha_i=1/2 [1-\tanh(1-\lambda_i/\overline{\lambda})]$.

Fig. \ref{fig:dyn} reports the corresponding scans of the electronic structure
at the Fermi energy (a) and at an energy of $0.6t$ below $E_F$.
Clearly the low energy electronic structure resembles that of a
homogeneous system due to the fact that the Ansatz Eq. \ref{eqs3} leads
to  $\Sigma_i(\omega=0)=0$.
Instead at high energies $\omega \gg \omega_0$ the scattering 
is controlled by $\Sigma_i(\omega \to \infty)= \Delta_i$ which
according to the above definitions is governed by the 
local charge and spin structure.
One can clearly see the similarities between the electronic structure 
at $\omega=-0.6t$ (cf. Fig. \ref{fig:dyn}b) and the reconstructed
Fermi surface from the static stripe solution shown in Fig. \ref{fig:fs}a.

Therefore this phenomenological theory reproduces our scenario
of low energy protected quasiparticles but emerging spectral
properties of charge and spin order at large energies. In contrast
to the Kampf-Schrieffer approach \cite{KAMPF} which is based on homogeneous
ground states the present theory explicitely describes
systems with broken symmetry, however, this broken symmetry manifests only at
high energies.

\section{Conclusions}\label{sec7}
In this paper we have reviewed evidence for stripe correlations
in cuprate superconductors based on Gutzwiller variational
calculations supplemented with Gaussian fluctuations.
Because our aim is to provide also a quantitative analyis
of static and dynamical properties, the appropriate parameter
set for lanthanum cuprates based on the extended Hubbard model 
was derived in Sec. \ref{sec1} in some detail. Our value of 
$U/t=8$ agrees with estimates from Ref. \cite{col01}
and the next-nearest neighbor hopping $t'/t=-0.2$ is
in the range of values as derived from LDA computations in
Ref. \cite{pav01}. 

The main emphasis of this review is put on the spin excitations
on top of stripe ground states. As we have discussed in Sec. \ref{sec4}
the time-dependent Gutzwiller approach yields an excellent
quantitative agreement with the spectra observed by 
INS when the stripes are static, i.e. in the LBCO compounds.
Upon underdoping and in the non-codoped system our calculations
in general overestimate the excitation energies. We attribute
this discrepancy to the much reduced correlation length as
extracted from the low energy spectra due to the dynamics
of the stripes and disorder effects \cite{sei09}.
In this regard it would also be interesting to compute the
spin excitations on top of the dynamical stripe solutions
which have been discussed in Sec. \ref{sec6}.

As one approaches the overdoped regime the nature of stripe correlations
in the ground state is not so clear 
since one has to take into account the melting of the stripes 
and non-Gaussian fluctuation effects will become important. 
Also it is conceivable that          
phase separation between the underdoped stripes and an overdoped Fermi 
liquid occurs as discussed in Ref.~\cite{lor02}. 
Evidence for this scenario was presented in Sec. \ref{sec4} from
the doping dependence of the saddle point energy.
Due to the fact that in the phase separation scenario the population     
of the domain walls by holes saturates a further           
increase of the saddle point energy with doping is hampered in 
agreement with the INS experiments in Ref.~\cite{vignolle}.  On the other
hand, Ref.~\cite{wak07} is in accord with a substantial increase of the
saddle point energy.

Another possiblity is that the anharmonic structure of the stripes is
gradually lost and stripes become harmonic dynamical and damped charge-density wave 
fluctuations. This leads to the more standard scenario of a second order
transition ending into a quantum critical point slightly above optimal doping\cite{cast95}
driven by a frustrated phase separation mechanism\cite{FPS}. In this case
the combined effect of charge and spin fluctuations can simply be treated
within perturbation theory\cite{sergio11}.

The natural question which arises is whether the results of Sec. \ref{sec4}
may then also be relevant for an understanding of spin        
excitations in YBCO. In this compounds the acoustic dispersion 
shows a low-energy spin gap
($\Delta \sim 30$meV at optimal doping)                                 
and thus does not reach the incommensurate wave-vector at $\omega=0$.   
Therefore it is more likely that the system is in a quantum          
disordered spin phase as suggested by the ladder                        
theories \cite{voj04,uhr04,uhr05}. Alternatively a scenario             
of fluctuating stripes where also the                                   
charge loses its long-range order can capture the effect of a           
spin-gap.\cite{vojta06}                                                   
One expects that the systems show short range order with               
a correlation length of the order of $\Delta/(\hbar c)\sim 5 a$         
and that for energies larger than $\Delta$ the                          
system resembles that of an ordered phase.                              
For this reason one can expect that a
computation as the one discussed in Sec. \ref{sec4} is suitable for a 
description of the     
universal high energy spin response  
\cite{tra04,chr04,vignolle,kofu07,matsuda08,matsuda11,hayden04,reznik04,stock05,hinkov08}.           
Again, to capture the excitations over the full energy range
would require an approach which incorporates the dynamical 
nature of stripes as the one proposed in Sec. \ref{sec6}.

Concluding, we have shown that the analyis of numerous experiments provides
strong evidence for stripe correlations in high-T$_c$ superconductors.
This more or less hidden electronic order can account  for both 
normal-state anomalies and high-temperature superconductivity in the
cuprates. 
In fact, the proximity to an instability (particularly if it is a 
second-order ``critical line'' ending at zero temperature into a quantum 
critical point) marking the onset of order naturally brings along abundant 
fluctuations and leads to strongly temperature- and doping-dependent 
features, which account for the non-Fermi liquid properties and for a
strong pairing interaction. The participation of such stripe fluctuations
in the 'pairing glue' is suggested by the recent analysis of Raman scattering
data \cite{sergio11} but in any case additional work has to be done in 
order to elucidate the source of 
strong scattering/pairing between the charge carriers leading to high-T$_c$.


\begin{thebibliography}{9}
\bibitem{bemu} J. G. Bednorz and K. A. M\"uller, Z. Physik B {\bf 64},
               189 (1986).
\bibitem{ps2} Proceedings of the {\it Second workshop on phase separation
              in cuprate superconductors}, E. Sigmund and K. A. M\"uller
              (eds.); Springer Verlag, Berlin Heidelberg (1994).
\bibitem{cast95} C. Castellani, C. Di Castro and M .Grilli, Phys. Rev. Lett.
                 {\bf 75}, 4650 (1995).
\bibitem{FPS}R. Raimondi, et al., Phys. Rev. B {\bf 47}, 3331 (1993); 
             V. J. Emery and S. Kivelson, Physica C {\bf 209}, 597 (1993);
            U. L\"ow et al., Phys. Rev. Lett. {\bf 72}, 1918 (1994); 
            Z. Nussinov et al., Phys. Rev. Lett. {\bf 83}, 472 (1999); 
            J. Lorenzana, C. Castellani, and C. Di Castro, 
            Phys. Rev. B {\bf 64}, 235127 (2001); 
            J. Lorenzana, C. Castellani, and C. Di Castro, 
            Europhys. Lett. {\bf 57}, 704 (2002); 
            R. Jamei, S. Kivelson, and B. Spivak, 
            Phys. Rev. Lett. {\bf 94}, 056805, (2005); 
            C. Ortix, J. Lorenzana, M. Beccaria, and C. Di Castro, 
            Phys. Rev. B {\bf 75}, 195107 (2007);  
            C. Ortix, J. Lorenzana, C. Di Castro, 
            Phys. Rev. Lett. 100. 246402 (2008).
\bibitem{zaa89} J. Zaanen and O. Gunnarsson, Phys. Rev. B {\bf 40}, 7391 (1989).
\bibitem{mac89} K. Machida, Physica C {\bf 158}, 192 (1989).
\bibitem{hsch90} H.~J. Schulz, Phys. Rev. Lett. {\bf 64}, 1445 (1990).
\bibitem{pol89} D. Poilblanc and T. M. Rice, Phys. Rev. B {\bf 39}, 9749 (1989).
\bibitem{WS} S. R. White and D. J. Scalapino, Phys. Rev. Lett. {\bf 80}, 1272
             (1998); S. R. White and D. J. Scalapino, Phys. Rev. Lett.
              {\bf 81}, 3227 (1998).
\bibitem{WS2} S. R. White and D. J. Scalapino, Phys. Rev. Lett. {\bf 91}, 
              136403, (2003).
\bibitem{WS3} S. R. White and D. J. Scalapino, Phys. Rev. B {\bf 60}, 
              R753 (1999)
\bibitem{HELLBERG} C. Stephen Hellberg and E. Manousakis, Phys. Rev. Lett.
                   {\bf 83}, 132 (1999).
\bibitem{TOHYAMA} T. Tohyama, C. Gazza, C. T. Shih, Y. C. Chen, T. K. Lee,
                  S. Maekawa, and E. Dagotto, Phys. Rev. B {\bf 59}, 
                  R11649 (1999).
\bibitem{BECCA} F. Becca, L. Capriotti, and S. Sorella, 
                Phys. Rev. Lett. {\bf 87}, 167005 (2001).        
\bibitem{HIMEDA} A. Himeda, T. Kato, and M. Ogata, Phys. Rev. Lett. {\bf 88},
                 117001 (2002).
\bibitem{fle01b} M. Fleck and A. I. Lichtenstein and E. Pavarini, Phys.
                 Rev. Lett. {\bf 84}, 4962 (2000);
                 M. Fleck and A. I. Lichtenstein and A. M. Ole\'s,
                 Phys. Rev. B {\bf 64}, 134528 (2001).
\bibitem{ichi99} K. Machida and M. Ichioka, J. Phys. Soc. Jpn. {\bf 68},
                 2168 (1999).
\bibitem{val00} B. Valenzuela, M. A. H. Vozmediano, and
                F. Guinea, Phys. Rev. B {\bf 62}, 11312 (2000).
\bibitem{norm02} B. Normand and A. P. Kampf, Phys. Rev. B {\bf 65},
                 020509 (2001).
\bibitem{ZACHER} M. G. Zacher, R. Eder, E. Arrigoni, and W. Hanke,
                 Phys. Rev. Lett. {\bf 85}, 824 (2000).
\bibitem{anisimov04} V. I. Anisimov, M. A. Korotin, A. S. Mylnikova, 
                 A. V. Kozhevnikov, Dm. M. Korotin, and J. Lorenzana, 
                 Phys. Rev. B {\bf 70}, 17250 (2004).  
\bibitem{sei981} G. Seibold, C. Castellani, C. Di Castro, and M. Grilli, 
                 Phys. Rev. B {\bf 58}, 13506 (1998).
\bibitem{sei982} G. Seibold, Phys. Rev. {\bf 58}, 15520 (1998).
\bibitem{wak07}S. Wakimoto, K. Yamada, J. M. Tranquada, C. D. Frost,
  R. J. Birgeneau, and H. Zhang, Phys. Rev. Lett. {\bf 98}, 247003 (2007).
\bibitem{lor02} J. Lorenzana and G. Seibold, Phys. Rev. Lett. {\bf 89}, 
                136401 (2002).
\bibitem{lor03}  J. Lorenzana and G. Seibold, Phys. Rev. Lett. {\bf 90}, 
                 066404 (2003).
\bibitem{sei04} G. Seibold and J. Lorenzana, Phys. Rev. B {\bf 69}, 
                134513 (2004).
\bibitem{sei07} G. Seibold, J. Lorenzana, and M. Grilli, 
                Phys. Rev. B {\bf 75}, 100505 (2007).
\bibitem{sei09} G. Seibold and J. Lorenzana, Phys. Rev. B {\bf 80}, 
                012509 (2009).
\bibitem{sei11} G. Seibold, R. S. Markiewicz, and J. Lorenzana, 
                Phys. Rev. B {\bf 83}, 205108 (2011).
\bibitem{tra95} J.~M. Tranquada,  B. J. Sternlieb, J. D. Axe, Y. Nakamura  
                and S. Uchida, Nature {\bf 375}, 56 (1995).   
\bibitem{tra96} J. M. Tranquada, J. D. Axe, N. Ichikawa, Y. Nakamura,
                S. Uchida, and B. Nachumi, Phys. Rev. B {\bf 54}, 7489 (1996). 
\bibitem{tra97} J. M. Tranquada, J. D. Axe, N. Ichikawa, A. R. Moodenbaugh, 
                Y. Nakamura, and S. Uchida, Phys. Rev. Lett. {\bf 78}, 
                338 (1997). 
\bibitem{nio1} S. M. Hayden, G. H. Lander, J. Zarestky, P. J. Brown,
               C. Stassis, P. Metcalf, and J. M. Honig,
               Phys. Rev. Lett. {\bf 68}, 1061 (1992).
\bibitem{nio2} C. H. Chen, S-W. Cheong and A. S. Cooper, Phys. Rev. Lett. {\bf
               71}, 2461 (1993).
\bibitem{nio3} J. M. Tranquada, D. J. Buttrey, V. Sachan, and  J. E. Lorenzo,
               Phys. Rev. Lett. {\bf 73}, 1003 (1994).
\bibitem{fujita021} M. Fujita, H. Goka, K. Yamada, and M. Matsuda,
                   Phys. Rev. Lett. {\bf 88}, 167008 (2002).
\bibitem{klauss00} Klauss, H.-H., Wagener, W., Hillberg, M., Kopmann, W.,
                  Walf, H., Litterst, F. J., H\"ucker, M.  and B\"uchner, B., 
                  Phys. Rev. Lett. {\bf 85}, 4590 (2000). 
\bibitem{huecker10} M. H\"ucker, M. v. Zimmermann, M. Debessai, 
                    J. S. Schilling, J. M. Tranquada, and G. D. Gu, 
                    Phys. Rev. Lett. {\bf 104}, 057004 (2010).  
\bibitem{fujita09}M. Fujita, M. Enoki, S. Iikubo, K. Kudo, N. Kobayashi, and 
                  K. Yamada, arXiv:0903.5391    
\bibitem{zimmermann} M. v. Zimmermann, A. Vigliante, T. Niem\"oller, 
                     N. Ichikawa, T. Frello, J. Madsen, P. Wochner, 
                     S. Uchida, N. H. Andersen, J. M. Tranquada, 
                     D. Gibbs and J. R. Schneider, EPL {\bf 41}, 629 (1998).
\bibitem{abb05} P. Abbamonte, A. Rusydi, S. Smadici, G.~D. Gu, 
                G.~A. Sawatzky, and D.~L. Feng, 
                Nature Phys. {\bf 1},  155  (2005).
\bibitem{fink09}  J. Fink, E. Schierle, E. Weschke, J.Geck, D. Hawthorn, 
                  V. Soltwisch, H. Wadati, and Hsueh-Hung Wu, 
                  Phys. Rev. B {\bf 79}, 100502 (2009).
\bibitem{yam98} K. Yamada, C. H. Lee, K. Kurahashi, J. Wada, S. Wakimoto,
                S. Ueki, H. Kimura, Y. Endoh, S. Hosoya, G. Shirane,
                R. J. Birgeneau, M. Greven, M. A. Kastner, and Y. J. Kim,
                Phys. Rev. B {\bf 57}, 6165 (1998).
\bibitem{waki04} S. Wakimoto, H. Zhang, K. Yamada, I. Swainson, Hyunkyung Kim, 
                 and R. J. Birgeneau, Phys.\ Rev.\ Lett. {\bf 92},  
                 217004  (2004). 
\bibitem{waki99} S. Wakimoto, G. Shirane, Y. Endoh, K. Hirota, S. Ueki,
K. Yamada, R. J. Birgeneau, M. A. Kastner, Y. S. Lee, P. M. Gehring and 
S. H. Lee, Phys. Rev. B 60, R769 (1999).
\bibitem{waki00} S. Wakimoto, R. J. Birgeneau, M. A. Kastner, Y. S. Lee, 
R. Erwin, P. M. Gehring, S. H. Lee, M. Fujita, K. Yamada, Y. Endoh, K. Hirota,
and G. Shirane, Phys. Rev. B {\bf 61}, 3699 (2000).
\bibitem{mats00} M. Matsuda, M. Fujita, K. Yamada, R. J. Birgeneau,
M. A. Kastner, H. Hiraka, Y. Endoh, S. Wakimoto, and G. Shirane,
Phys. Rev. B {\bf 62}, 9148 (2000).
\bibitem{fujita02} M. Fujita, K. Yamada, H. Hiraka, P. M. Gehring,
S. H. Lee, S. Wakimoto and G. Shirane, Phys. Rev. B {\bf 65}, 064505 (2002).
\bibitem{fujita04}
M. Fujita, H. Goka, K. Yamada, J.~M. Tranquada, and L.~P. Regnault, Phys.\
  Rev.\ B {\bf 70},  104517  (2004).                                      
\bibitem{matsuda02} M. Matsuda, M. Fujita, K. Yamada, R. J. Birgeneau,
Y. Endoh, and G. Shirane, Phys. Rev. B {\bf 65}, 134515 (2002). 
\bibitem{arai99} M. Arai, , T. Nishijima, Y. Endoh, T. Egami, S. Tajima, 
                 K. Tomimoto, Y. Shiohara, M. Takahashi, A. Garrett,  
                 and S. M. Bennington, Phys. Rev. Lett. {\bf 83}, 608 (1999). 
\bibitem{dai01} P. Dai, H. A. Mook, R. D. Hunt, and F. Dogan, 
                Phys. Rev. B {\bf 63}, 054525 (2001).
\bibitem{haug10} D. Haug, V. Hinkov, Y. Sidis, P. Bourges, N. B. Christensen, 
                 A. Ivanov, T Keller, C. T. Lin and B. Keimer, New J. Phys. 
                 {\bf 12}, 105006 (2010).
\bibitem{note1} In some measurements only an approach of incommensurate 
                branches towards Q$_{AF}$ is resolved with a subsequent 
                dispersion outwards with increasing 
                energy.
\bibitem{tra04}
J.~M. Tranquada, H. Woo, T.~G. Perring, H. Goka, G.~D. Gu, G. Xu, M. Fujita,
  and K. Yamada, Nature (London) {\bf 429},  534  (2004).                   
\bibitem{hiraka01} H. Hiraka {\it et al.}, J. Phys. Soc. Jpn. {\bf 70},
                    853 (2001).
\bibitem{chr04}
N.~B. Christensen, D.~F. McMorrow, H.~M. R{\o}nnow, B. Lake, S.~M. Hayden, G.
  Aeppli, T.~G. Perring, M. Mangkorntong, M. Nohara, and H. Tagaki, Phys.\   
  Rev.\ Lett. {\bf 93},  147002  (2004). 
\bibitem{vignolle} B. Vignolle, S. M. Hayden, D. F. McMorrow, H. M. Ronnow, C.D. Frost, and T. G. Perring, Nat. Phys. {\bf 3}, 163 (2007).
\bibitem{kofu07} M. Kofu, T. Yokoo, F. Trouw, and K. Yamada,arXiV:0710:5766.
\bibitem{hin04}
V. Hinkov, S. Pailh{\'e}s, P. Bourges, Y. Sidis, A. Ivanov, A. Kulakov, C. Lin,
  D. Chen, C. Bernhard, and B. Keimer, Nature (London) {\bf 430},  650  (2004).
\bibitem{matsuda08} M. Matsuda, M. Fujita, S. Wakimoto,
  J. A. Fernandez-Baca, J. M. Tranquada, and K. Yamada,
  Phys. Rev. Lett. {\bf 101}, 197001 (2008).
\bibitem{sca98} D. J. Scalapino and  S. R. White, Phys. Rev. B {\bf
    58}, 8222 (1998).   
\bibitem{matsuda11} M. Matsuda, J. A. Fernandez-Baca, M. Fujita, K. Yamada, and J. M. Tranquada, Phys. Rev. B {\bf 84}, 104524 (2011). 
\bibitem{hayden04} S. M. Hayden, H. A. Mook, Pengcheng Dai, T. G. Perring and
                   F. Dogan, Nature {\bf 429}, 531 (2004).
\bibitem{reznik04} D. Reznik, P. Bourges, L. Pintschovius, Y. Endoh, 
                   Y. Sidis, T. Masui, and S. Tajima, Phys. Rev. Lett. {\bf 93},
                   207003 (2004).
\bibitem{stock05} C. Stock, W. J. L. Buyers, R. A. Cowley, P. S. Clegg, 
                  R. Coldea, C. D. Frost, R. Liang, D. Peets, D. Bonn, 
                  W. N. Hardy, and R. J. Birgeneau, Phys. Rev. B {\bf 71}, 
                  024522 (2005).
\bibitem{hinkov07} Hinkov, P. Bourges, S. Pailh\'{e}s, Y. Sidis, A. Ivanov, 
                  C. D. Frost, T. G. Perring, C. T. Lin, D. P. Chen, and  
                  B. Keimer, Nat. Phys. {\bf 3}, 780 (2007).
\bibitem{hinkov08} V. Hinkov, D. Haug, B. Fauqu\'{e}, P. Bourges, Y. Sidis, 
                   A. Ivanov, C. Bernhard, C. T. Lin and B. Keimer, 
                  Science {\bf 319} (2008).
\bibitem{fauque07}  B. Fauqu\'{e}, Y. Sidis, L. Capogna, A. Ivanov, K. Hradil, 
                 C. Ulrich, A. I. Rykov, B. Keimer, and P. Bourges, 
                 Phys. Rev. B {\bf 76}, 214512 (2007).  
\bibitem{xu09} Guangyong Xu, G. D. Gu, M. H\"ucker, B. Fauqu\'{e}, T. G. 
               Perring, L. P. Regnault, and J. M. Tranquada, 
               Nat. Phys. {\bf 5}, 642 (2009).
\bibitem{sei05}
G. Seibold and J. Lorenzana, Phys.\ Rev.\ Lett. {\bf 94},  107006  (2005).
\bibitem{sei06} G. Seibold and J. Lorenzana, Phys. Rev. B{\bf 73}, 
144515 (2006).
\bibitem{voj04}
M. Vojta and T. Ulbricht, Phys.\ Rev.\ Lett. {\bf 93},  127002  (2004).
\bibitem{uhr04}
G.~S. Uhrig, K.~P. Schmidt, and M. {Gr\"uninger}, Phys.\ Rev.\ Lett. {\bf 93},
  267003  (2004).                                                             
\bibitem{uhr05}
G.~S. Uhrig, K.~P. Schmidt, and M. {Gr\"uninger}, J.Phys.Soc.Jpn. {\bf 74} (2005), Supplement 86.
\bibitem{moe04}
B.~M. Andersen and P. Hedeg{\aa}rd, Phys.\ Rev.\ Lett. {\bf 95},  037002
  (2005).
\bibitem{vojta06} M. Vojta, T. Vojta, and R. K. Kaul, Phys. Rev. Lett. {\bf 97}, 097001 (2007). 
\bibitem{and07} Brian M. Andersen and Olav F. Sylju\aa{}sen, Phys. Rev. B 
{\bf 75}, 012506 (2007).  
\bibitem{and10} Brian M. Andersen, Siegfried Graser, and P. J. Hirschfeld, Phys. Rev. Lett. {\bf 105}, 147002 (2010).   
\bibitem{kru03}
F. Kr\"uger and S. Scheidl, Phys.\ Rev.\ B {\bf 67},  134512  (2003).
\bibitem{car04}
E.~W. Carlson, D.~X. Yao, and D.~K. Campbell, Phys.\ Rev.\ B {\bf 70},  064505
  (2004).
\bibitem{car06} D.~X. Yao, E.~W. Carlson,  and D.~K. Campbell, Phys. Rev. Lett. {\bf 97}, 017003 (2006).
\bibitem{boothroyd11} A. T. Boothroyd, P. Babkevich, D. Prabhakaran, and 
P. G. Freeman, Nature {\bf 471}, 341 (2011).
\bibitem{kraem99} S. Kr\"amer and M. Mehring, Phys. Rev. Lett. {\bf 83}, 
                  396 (1999).
\bibitem{tei00} G. B. Teitel'baum, B. B\"uchner, and H. de Gronckel,
                Phys. Rev. Lett. {\bf 84}, 2949 (2000).
\bibitem{sing02} P. M. Singer, A. W. Hunt, and T. Imai, Phys. Rev. Lett.
                 {\bf 88}, 047602 (2002).
\bibitem{has02} J. Haase, C. P. Slichter, and C. T. Milling, 
                J. Supercond. {\bf 15}, 339 (2002).
\bibitem{julien11} T. Wu, H. Mayaffre, S. Kramer, M. Horvatic, C. Berthier, W.N. Hardy, R. Liang, D.A. Bonn, and M.-H. Julien, Nature {\bf 477}, 191 (2011).
\bibitem{vojta} M. Vojta, Adv. Phys. {\bf 58}, 699 (2009).
\bibitem{daou10} R. Daou, J. Chang, D. LeBoeuf, O. Cyr-Choiniere, 
                F. Laliberte, N. Doiron-Leyraud, B. J. Ramshaw, R. Liang, 
                D. A. Bonn, W. N. Hardy, and L. Taillefer, 
                Nature {\bf 463}, 519 (2010).
\bibitem{brinckmann99}, Jan Brinckmann and Patrick A. Lee, Phys. Rev. Lett. 
                        {\bf 82}, 2915 (1999).
\bibitem{norman00} M. R. Norman,Phys. Rev. B {\bf 61}, 14751 (2000).
\bibitem{chubu01}   Andrey V. Chubukov, Boldizs\'{a}r Jank\'{o}, and 
                    Oleg Tchernyshyov, Phys. Rev. B {\bf 63}, 180507(R) (2001). \bibitem{schnyder04} A. P. Schnyder, A. Bill, C. Mudry, R. Gilardi, 
                     H. M. Ronnow, and J. Mesot, 
                     Phys. Rev. B {\bf 70}, 214511 (2004).     
\bibitem{berciu04} M. Berciu and S. John, Phys. Rev. B {\bf 69}, 224515 (2004).
\bibitem{sherman04} A. Sherman and M. Schreiber, Phys. Rev. B {\bf 69}, 
                    100505 (2004).
\bibitem{eremin05} I. Eremin, D. K. Morr, A. V. Chubukov, K. H. Bennemann, 
                   and M. R. Norman,  
                   Phys. Rev. Lett. {\bf 94}, 147001 (2005).  
\bibitem{sherman11} A. Sherman, arXiv:1108.4179.
\bibitem{yamase06} Hiroyuki Yamase and Walter Metzner, Phys. Rev. B {\bf 73}, 214517 (2006).  
\bibitem{reznik10} D. Reznik, Adv. Cond. Matt. Phys. {\bf 2010}, 
                   Article ID 523549. 
\bibitem{kane02} E. Kaneshita, M. Ichioka, and K.Machida, Phys.
                 Rev. Lett. {\bf 88}, 115501 (2002).
\bibitem{mukhin07} S. I. Mukhin, A. Mesaros, J. Zaanen, and F. V. Kusmartsev,
                 Phys. Rev. B {\bf 76}, 174521 (2007).
\bibitem{salkola} M.I. Salkola, V.J. Emery, and S.A. Kivelson, 
                  J. Supercond. {\bf 9}, 401 (1996).
\bibitem{sei00} G. Seibold, F. Becca, F. Bucci, C. Castellani, C. Di Castro, 
M. Grilli, Eur. Phys. J B {\bf 13}, 87 (2000).
\bibitem{zho99} X. J. Zhou, P. Bogdanov, S. Kellar, T. Noda, H. Eisaki,
               S. Uchida, Z. Hussain, and Z.-X. Shen, Science {\bf 286},
               268 (1999).
\bibitem{zho01}  X.~J. Zhou, T. Yoshida, S.~A. Kellar, P.~V. Bogdanov, 
                 E.~D. Lu, A. Lanzara, M. Nakamura, T. Noda, 
                 T. Kakeshita, H. Eisaki, S. Uchida, A. Fujimori, Z. Hussain, 
                 and Z.~X. Shen, Phys. Rev. Lett. {\bf 86},  5578  (2001).
\bibitem{ino97} A. Ino, T. Mizokawa, A. Fujimori, K. Tamasaku, H. Eisaki,
                S. Uchida, T. Kimura, T. Sasagawa, and K. Kishio,
                Phys. Rev. Lett. {\bf 79}, 2101 (1997).
\bibitem{har01} N. Harima, J. Matsuno, A. Fujimori, Y. Onose, Y. Taguchi,
                and Y. Tokura, Phys. Rev. B {\bf 64}, R220507 (2001). 
\bibitem{doiron1} N. Doiron-Leyraud, C. Proust, D. LeBoeuf, 
                  J. Levallois, J.-B. Bonnemaison, R. Liang, D. A. Bonn, 
                  W. N. Hardy, and L. Taillefer, Nature {\bf 447}, 565 (2007).
\bibitem{yelland} E. A. Yelland, J. Singleton, C. H. Mielke, 
                    N. Harrison, F. F. Balakirev, B. Dabrowski, and 
                    J. R. Cooper, arXiv:0707.0057.
\bibitem{bangura} A. F. Bangura, J. D. Fletcher, A. Carrington, 
                   J. Levallois, M. Nardone, B. Vignolle, P. J. Heard, 
                   N. Doiron-Leyraud, D. LeBoeuf, L. Taillefer, S. Adachi, 
                   C. Proust, and N. E. Hussey, arXiv:0707.4461.
\bibitem{leboeuf} D. LeBoeuf, N. Doiron-Leyraud, R. Daou, J.-B. Bonnemaison, 
                  J. Levallois, N. E. Hussey, C. Proust, L. Balicas, 
                  B. Ramshaw, R. Liang, D. A. Bonn, W. N. Hardy, 
                  S. Adachi, and L. Taillefer, Nature {\bf 450}, 533 (2007).
\bibitem{millis07} A. J. Millis and M. R. Norman, Phys. Rev. B {\bf 76}, 220503(R) (2007).
\bibitem{laliberte11} F. Laliberte, J. Chang, N. Doiron-Leyraud, E. Hassinger, R. Daou, M. Rondeau, B. J. Ramshaw, R. Liang, D. A. Bonn, W. N. Hardy, S. Pyon, T. Takayama, H. Takagi, I. Sheikin, L. Malone, C. Proust, K. Behnia, 
L. Taillefer, Nature Communications {\bf 2}, 432 (2011).
\bibitem{wang06} Wang, Y., Li, P., and Ong, N. P., Phys. Rev. B {\bf 73}, 
024510 (2006). 
\bibitem{choin09} Olivier Cyr-Choinière, R. Daou, Francis Lalibert\'{e}, 
David LeBoeuf, Nicolas Doiron-Leyraud, J. Chang, J.-Q. Yan, J.-G. Cheng, 
J.-S. Zhou, J. B. Goodenough, S. Pyon, T. Takayama, H. Takagi, Y. Tanaka, and  
Louis Taillefer, Nature {\bf 458}, 743 (2009).
\bibitem{hess10} Christian Hess, Emad M. Ahmed, Udo Ammerahl, Alexandre 
                 Revcolevschi, Bernd B\"uchner, Eur. Phys. 
                 J. Special Topics {\bf 188}, 103 (2010).
\bibitem{martin10}  Ivar Martin and C. Panagopoulos, Eur. Phys. Lett.  
{\bf 91}, 67001 (2010).
\bibitem{hackl10} Andreas Hackl and Matthias Vojta, New J. Phys. {\bf 12}, 
105011 (2010).
\bibitem{hoff02} J. E. Hoffmann, K. McElroy, D.-H. Lee, K. M Lang, H. Eisaki, 
                 S. Uchida and J. C. Davis, Science {\bf 297}, 1148 (2002).
\bibitem{how03} C. Howald, H. Eisaki, N. Kaneko, M. Greven, and 
                A. Kapitulnik, Phys. Rev. B {\bf 67}, 014533 (2003).
\bibitem{elroy05} K. McElroy, D.-H. Lee, J. E. Hoffman, K. M. Lang, J. Lee, 
                  E. W. Hudson, H. Eisaki, S. Uchida, and J. C. Davis,   
                  Phys. Rev. Lett. {\bf 94}, 197005 (2005).
\bibitem{hashi06} A. Hashimoto, N. Momono, M. Oda, and M. Ido, 
                  Phys. Rev. B {\bf 74}, 064508 (2006).
\bibitem{hana07} T. Hanaguri, Y. Kohsaka, J. C. Davis, C. Lupien, 
                 I. Yamada, M. Azuma, M. Takano, K. Ohishi, M. Ono, and
                 H. Takagi, Nature Physics {\bf 3}, 865 (2007).
\bibitem{wise08} W. D. Wise, M. C. Boyer, Kamalesh Chatterjee, 
                 Takeshi Kondo, T. Takeuchi, H. Ikuta, Yayu Wang,  and  
                 E. W. Hudson, Nature Physics {\bf 4}, 696 (2008).
\bibitem{versh04} M. Vershinin, Shashank Misra1, S. Ono, Y. Abe2, 
                  Yoichi Ando and Ali Yazdani, Science {\bf 303}, 1995 (2004). 
\bibitem{hana04} T. Hanaguri, C. Lupien, Y. Kohsaka, D.-H. Lee, M. Azuma, 
                 M. Takano, H. Takagi, and J. C. Davis, 
                 Nature {\bf 430}, 1001 (2004). 
\bibitem{wang03} Q. H. Wang and D.-H. Lee, Phys. Rev. B {\bf 67}, 
                 020511(R) (2003).
\bibitem{kohsaka08} Y. Kohsaka, C. Taylor, P. Wahl, A. Schmidt, Jhinhwan Lee, 
                    K. Fujita, J. W. Alldredge, K. McElroy, Jinho Lee, 
                    H. Eisaki, S. Uchida, D.-H. Lee, and  J. C. Davis, 
                    Nature {\bf 454}, 1072 (2008).
\bibitem{alldredge08} J. W. Alldredge, Jinho Lee, K. McElroy, M. Wang, 
                     K. Fujita, Y. Kohsaka, C. Taylor, H. Eisaki, S. Uchida, 
                     P. J. Hirschfeld, and J. C. Davis, 
                     Nature Physics {\bf 4}, 319 (2008).
\bibitem{gut65} M.~C. Gutzwiller, Phys. Rev. {\bf 137}, A1726 (1965).
\bibitem{sei01} G. Seibold and J. Lorenzana, Phys. Rev. Lett. {\bf 86}, 
                2605 (2001).
\bibitem{goe03} G. Seibold, F. Becca, and J. Lorenzana, Phys. Rev. B {\bf 67},
    085108 (2003).
\bibitem{goe042} G. Seibold, F. Becca, P. Rubin, and
                 J. Lorenzana, Phys. Rev. B {\bf 69}, 155113 (2004).
\bibitem{goe08} G. Seibold, F. Becca, and J. Lorenzana,
                Phys. Rev. Lett. {\bf 100}, 016405 (2008);
                G. Seibold, F. Becca, and J. Lorenzana,
                Phys. Rev. B {\bf 78}, 045114 (2008).
\bibitem{ichi01} M. Ichioka, E. Kaneshita, and K. Machida, J. Phys. Soc. Jpn.
                {\bf 70}, 818 (2001).
\bibitem{kan01} E. Kaneshita, M Ichioka, and K. Machida
  J. Phys. Soc. Jpn, {\bf 70}, 866 (2001). 
\bibitem{kan02} E. Kaneshita, R. Morino, M. Ichioka, and K. Machida, J.
                 Phys. Chem. Solids {\bf 63}, 1545 (2002).
\bibitem{kan022} E. Kaneshita, M. Ichioka, and K. Machida, Phys. Rev. Lett. 
                 {\bf 88}, 115501 (2002).
\bibitem{varl02} S. Varlamov and G. Seibold, Phys. Rev. B {\bf 65}, 
                 075109 (2002).
\bibitem{inu91} M. Inui and P. B. Littlewood, Phys. Rev. B {\bf 44},
                4415 (1991).
\bibitem{singh95}  R. R. P. Singh and M. P. Gelfand, Phys. Rev. 
                 B {\bf 52}, R15695 (1995).
\bibitem{col01} R. Coldea, S. M. Hayden, G. Aeppli, T. G. Perring, C. D.
                  Frost, T. E. Mason, S.-W. Cheong, and Z. Fisk,
                  Phys. Rev. Lett. {\bf 86}, 5377 (2001).
\bibitem{headings10} N. S. Headings, S. M. Hayden, R. Coldea, and T. G. Perring
Phys. Rev. Lett. {\bf 105}, 247001 (2010). 
\bibitem{rog89} M. Roger and J.~M. Delrieu, Phys.\ Rev.\ B 
                {\bf 39},  2299  (1989).
\bibitem{sch90} H.~J. Schmidt and Y. Kuramoto, Physica C {\bf 167},  
                263  (1990).
\bibitem{lem97} F. Lema, J.~M. Eroles, C.~D. Batista, and E. Gagliano, 
                Phys.\ Rev.\ B {\bf 55}, 15295  (1997).
\bibitem{lor99} J. Lorenzana, J. Eroles, and S. Sorella, 
                Phys.\ Rev.\ Lett. {\bf 83},  5122 (1999).
\bibitem{zheng05} Weihong Zheng, Rajiv R. P. Singh, Jaan Oitmaa, 
Oleg P. Sushkov, and Chris J. Hamer, Phys. Rev. B {\bf 72}, 033107 (2005).
\bibitem{uch91} S. Uchida, T. Ido, H. Takagi, T. Arima, Y. Tokura,
                and S. Tajima, Phys. Rev. B {\bf 43}, 7942 (1991).
\bibitem{oka11} H. Okamoto, T. Miyagoe, K. Kobayashi, H. Uemura, H. Nishioka, H. Matsuzaki, A. Sawa, and Y. Tokura, Phys. Rev. B  {\bf 83}, 125102 (2011).
\bibitem{comnest} Strictly speaking this coincidence only shows up for
                  $n_h \gtrsim 0.1$ and susceptibilities evaluated with 
                  electronic structure parameters appropriate for LSCO.        
\bibitem{marki10} R. S. Markiewicz, J. Lorenzana, and G. Seibold, Phys. Rev. 
                 B {\bf 81}, 014510 (2010).
\bibitem{granath04} M. Granath, Phys. Rev. B {\bf 69}, 214433 (2004).
\bibitem{esk90} H.~Eskes, L.~H. Tjeng, G.~A. Sawatzky, Phys. Rev. B {\bf 42}, 
                 288 (1990).
\bibitem{koh07} Y. Kohsaka, C. Taylor, K. Fujita, A. Schmidt, C. Lupien, 
                T. Hanaguri, M. Azuma, M. Takano, H. Eisaki, H. Takagi, 
                S. Uchida, and J. C. Davis,  Science {\bf 315}, 1380 (2007).
\bibitem{lor05} J. Lorenzana, G. Seibold, and R. Coldea, Phys. Rev. B {\bf 72}
                224511 (2005).
\bibitem{normand01} B. Normand and A. P. Kampf, Phys. Rev. B {\bf 64}, 024521 (2001).
\bibitem{he10} Rui-Hua He, M. Fujita, M. Enoki, M. Hashimoto, S. Iikubo, S.-K. Mo, Hong Yao, T. Adachi, Y. Koike, Z. Hussain, Z.-X. Shen, and K. Yamada, Phys. Rev. Lett. {\bf 107}, 127002 (2011).  
\bibitem{rac06} M. Raczkowski, R. Fr\'{e}sard, and A. M. Ole\'{s},
                Europhysics Letters {\bf 76}, 128 (2006).
\bibitem{rac07} M. Raczkowski, R. Fr\'{e}sard, and A. M. Ole\'{s},
                Phys. Stat. Sol. (b) {\bf 244}, 2521 (2007). 
\bibitem{mar00} G. B. Martins, C. Gazza, J. C. Xavier, A. Feiguin,
                  and E. Dagotto, Phys. Rev. Lett. {\bf 84}, 5844 (2000).
\bibitem{brai10} L. Braicovich, J. van den Brink, V. Bisogni, M. Moretti 
                 Sala, L. J. P. Ament, N. B. Brookes, G. M. De Luca, 
                 M. Salluzzo, T. Schmitt, V. N. Strocov, and G. Ghiringhelli,
                 Phys. Rev. Lett. {\bf 104}, 077002 (2010). 
\bibitem{suz89} M. Suzuki, Phys. Rev. B {\bf 39}, 2312 (1989).                P
\bibitem{ter99} M. Terauchi and M. Tanaka, Micron {\bf 30},  371  (1999).
\bibitem{mcm90} A. K. McMahan, J. F. Annett, and R. M. Martin, Phys. Rev.
                B {\bf 42}, 6268 (1990).
\bibitem{hybert} M. S. Hybertsen, E. B. Stechel, W. M. C. Foulkes, and M. 
                 Schl\"uter, Phys. Rev. B {\bf 45}, 10032 (1992). 
\bibitem{man11}B. Mansart, J. Lorenzana, M. Scarongella, M. Chergui
  and F. Carbone,  arXiv:1112.0737.
\bibitem{homes03} C. C. Homes, J. M. Tranquada*, Q. Li, A. R. Moodenbaugh, 
                  D. J. Buttrey,  Phys. Rev. B {\bf 67}, 184516 (2003). 
\bibitem{sei01b} G. Seibold and M. Grilli, Phys. Rev. B {\bf 63},
                224505 (2001).
\bibitem{grilli05} G. Seibold and M. Grilli, Phys. Rev B {\bf 72}, 104519 
                  (2005).
\bibitem{grilli09} M. Grilli, G. Seibold, A. Di Ciolo, and J. Lorenzana,
                   Phys. Rev. B {\bf 79}, 125111 (2009).
\bibitem{sei10} G. Seibold, M. Grilli, and J. Lorenzana,
                New Journal of Physics {\bf 12}, 105010 (2010).
\bibitem{ma08} J.-H. Ma, Z.-H. Pan, F. C. Niestemski, M. Neupane, Y.-M. Xu, 
               P. Richard, K. Nakayama, T. Sato, T. Takahashi, H.-Q. Luo, 
               L. Fang, H.-H. Wen, Ziqiang Wang, H. Ding, and V. Madhavan, 
               Phys. Rev. Lett. {\bf 101}, 207002 (2008).
\bibitem{KAMPF} A. P. Kampf and J. R. Schrieffer, Phys. Rev. B{\bf 42},
                7967 (1990).
\bibitem{kondo09} Takeshi Kondo, Rustem Khasanov, Tsunehiro Takeuchi, 
                  J\"org Schmalian and Adam Kaminski, Nature {\bf 457},
                  296 (2009).
\bibitem{GWEON} G.-H. Gweon, T. Sasagawa, S.Y. Zhou, J. Graf, H. Takagi,
                D.-H. Lee, and A. Lanzara, Nature {\bf 430}, 187 (2004).
\bibitem{chubukov} Ar. Abanov, A. Chubukov, and J. Schmalian, Adv. Phys. 52, {\bf 119} (2003), 
and references therein.
\bibitem{ENSS}S. Caprara, M. Grilli, C. Di Castro, and T. Enss, Phys. Rev. B {\bf 75},
                140505 (2007).
\bibitem{CAPRARA}S. Caprara, C. Di Castro, B. Muschler, W. Prestel, R. Hackl,  M. Lambacher,
 A. Erb,  S. Komiya, Y. Ando, and M. Grilli, Phys. Rev. B {84}, 0504508 (2011).
 \bibitem{BIANCONI}A. Bianconi, N. L. Saini, A. Lanzara, M. Missori, T. Rossetti, H. Oyanagi, H. Yamaguchi, K. Oka, and T. Ito, Phys. Rev. Lett. {\bf 76}, 3412Ð3415 (1996).
 \bibitem{FRATINI}M. Fratini, N. Poccia, A. Ricci, G. Campi, M. Burghammer, G. Aeppli, and A. Bianconi,
 Nature {\bf 466}, 841 (2010).
\bibitem{sei092} G. Seibold, M. Grilli, and J. Lorenzana,
                 Phys. Rev. Lett. {\bf 103}, 217005 (2009).
\bibitem{pav01} E. Pavarini, I. Dasgupta, T. Saha-Dasgupta, 
                O. Jepsen, and O. K. Andersen  Phys. Rev. Lett. {\bf 87}, 
                047003 (2001).
\bibitem{sergio11}S. Caprara, C. Di Castro, B. Muschler, W. Prestel, R.
 Hackl, M. Lambacher, A. Erb, S. Komiya, Y. Ando, and M. Grilli,       
Phys. Rev. B {\bf 84}, 054508 (2011).
\end{thebibliography}
\end{document}